\pdfoutput=1

\documentclass[twocolumn,showpacs,preprintnumbers,amsmath,amssymb,superscriptaddress]{revtex4}

\usepackage{graphics}
\usepackage{epstopdf}
\usepackage{graphicx}
\usepackage{dcolumn}
\usepackage{bm}
\usepackage{amsmath}
\usepackage{hyperref}
\usepackage[]{color}

\newcommand{\be}{\begin{equation}}
\newcommand{\ee}{\end{equation}}
\newcommand{\beq}{\begin{eqnarray}}
\newcommand{\eeq}{\end{eqnarray}}

\def\nue{\mathrel{{\nu_e}}}
\def\numu{\mathrel{{\nu_\mu}}}
\def\nutau{\mathrel{{\nu_\tau}}}
\def\nux{\mathrel{{\nu_x}}}

\def\barnue{\mathrel{{\bar \nu}_e}}
\def\barnumu{\mathrel{{\bar \nu}_\mu}}
\def\barnutau{\mathrel{{\bar \nu}_\tau}}

\def \lta {\mathrel{\vcenter{\hbox{$<$}\nointerlineskip\hbox{$\sim$}}}}
\def \gta {\mathrel{\vcenter{\hbox{$>$}\nointerlineskip\hbox{$\sim$}}}}

\def\t13{\mathrel{{\theta_{13}}}}

\def\s213{\mathrel{{\sin^2 \theta_{13}}}}

\def\y12{\mathrel{{\tan^2 \theta_{12}}}}
\def\c2{\mathrel{{\chi^2 }}}

\newcommand{\n}{neutrino}
\newcommand{\ns}{neutrinos}

\newcommand{\sn}{supernova}
\newcommand{\sne}{supernovae}

\newcommand{\oss}{oscillations}
\newcommand{\vac}{vacuum}

\newcommand{\fe}{Fe-core}
\newcommand{\one}{ONeMg-core}

\begin{document}

\preprint{RBRC 712}

\title{Neutrino oscillation signatures of oxygen-neon-magnesium supernovae} 

\author{C. Lunardini}
\affiliation{Arizona State University, Tempe, AZ 85287-1504}%
\affiliation{RIKEN BNL Research Center, Brookhaven National Laboratory, Upton, NY 11973}

\author{B. M\"uller}
\affiliation{Max-Planck-Institut f\"ur Astrophysik, Karl-Schwarzschild-Str. 1, D-85741 Garching, Germany}%

\author{H.-Th. Janka}
\affiliation{Max-Planck-Institut f\"ur Astrophysik, Karl-Schwarzschild-Str. 1, D-85741 Garching, Germany}%

 

\begin{abstract}
  We discuss the flavor conversion of neutrinos from core collapse
  supernovae that have oxygen-neon-magnesium (ONeMg) cores. Using the
  numerically calculated evolution of the star up to 650 ms
  post bounce, we find that, for the normal mass hierarchy, the
  electron neutrino flux in a detector shows signatures of two typical
  features of an ONeMg-core supernova: a sharp step in the density profile
  at the base of the He shell and a faster shock wave propagation
  compared to iron core supernovae.  Before the shock hits the density
  step ($t \lta 150$ ms), the survival probability of electron
  neutrinos above $\sim$20 MeV of energy is about $\sim 0.68$, in contrast to values of $\sim
  0.32$ or less for an iron core supernova.  The passage of the shock
  through the step and its subsequent propagation cause a decrease of
  the survival probability and a decrease of the amplitude of
  oscillations in the Earth, reflecting the transition to a more
  adiabatic propagation inside the star.  These changes affect the
  lower energy neutrinos first; they are faster and more sizable for
  larger $\theta_{13}$.  They are unique of ONeMg-core supernovae, and give
  the possibility to test the speed of the shock wave.  The time
  modulation of the Earth effect and its negative sign at the
  neutronization peak are the most robust signatures in a
  detector. 
\end{abstract}

\maketitle
  
\section{Introduction}
\label{intro}

A core collapse \sn\ is mainly a \n\ phenomenon. The $\sim 3 \cdot 10^{53}$
ergs of energy liberated in the collapse are radiated by neutrino
emission from a thermal surface, the neutrinosphere.  

While the \n\ flux at the neutrinosphere is fairly independent of the
 star's properties, the \n\ flux we receive on Earth does depend on
 those.  Indeed, \n\ oscillations are sensitive to the profile of the
 matter density that the \ns\ encounter on their path. This  makes
 them a very valuable tool to do a tomography of the star, in a way that
 depends on the \n\ mass spectrum and mixing matrix.  An example is
 the possibility to use \ns\ to track the distortions in density
 caused by the propagation of the shock wave inside the star \cite{Schirato:2002tg}.
 The sensitivity of the shock effects to the mixing $\t13$ would
 provide a powerful test of it \cite{Takahashi:2002yj,Lunardini:2003eh,Fogli:2003dw,Tomas:2004gr,Fogli:2004ff}.

An  interesting implication of  \n\ \oss\ as tools for  star tomography is
the possibility to distinguish between different types of
supernovae that differ in their density profile.  In particular, \ns\ 
can distinguish  \cite{Duan:2007sh} between a \sn\ with an iron core (\fe\ \sn) and  one
with an oxygen-neon-magnesium core (\one\ \sn), for which the density
distribution and the shock propagation are completely different.

Stars in the mass window between roughly eight and ten solar masses 
are expected to develop ONeMg cores, which may undergo gravitational
collapse before Ne ignition due to rapid electron captures on 
$^{24}$Mg and $^{20}$Ne and may thus explode as so-called
electron-capture supernovae (see \cite{Poelarends:2007ip} for a recent study with a summary of
previous works). 
This distinguishes such stars from the more
massive supernova progenitors, whose evolution through all stages of 
hydrostatic nuclear burning leads to the formation of an iron core at 
their center. ONeMg cores are bounded by an extremely steep density
gradient (which we will call ``density step'' in the following),
which differs drastically from the much shallower density 
profiles around iron cores (see, e.g., figs.~1 of \cite{Janka:2007yu,Janka:2007di}).

The steep density decline allows the hydrodynamic bounce-shock to
expand continuously, a fact that is favorable for efficient neutrino 
heating in the post-shock layer and thus facilitates neutrino-driven
explosions even in spherically symmetric simulations \cite{WilsonMayle,Kitaura:2005bt}. Traveling down the density step, the outgoing 
shock also accelerates to much higher velocities than in the central
regions of Fe-core supernovae. Despite the large difficulties in modeling
their progenitor evolution, ONeMg core collapses have received quite
some interest, in particular because of repeated speculations that
they might be the long-sought site of the formation of r-process
elements (e.g., \cite{hil78,whecowhi98,sumter01,wanom03,ningqian}) and
because of their possible link to the supernova of 1054 A.D., which
gave birth to the Crab nebula (e.g., \cite{gott70,arn75,woo80,hil82,nom82}).
The number of similar stellar death events could account for a fair
fraction of all supernovae. Poelarends et al. \cite{Poelarends:2007ip} estimate that
about 4\% of the stellar core collapses in the local universe might
be of this kind, but the uncertainties in modeling the stellar
evolution in the initial mass range between 6 and 12$\,M_\odot$ are
significant and therefore the authors do not exclude a contribution
of even 20\% to all supernovae.

To identify or exclude the oscillation signatures of an \one\ \sn\ in an observed \n\ burst would be an important, unprecedented, test of stellar models and in particular of the existence of a density step in the interior of the star. It  would be a necessary part of any data analysis aimed at reconstructing the originally produced \n\ fluxes (before \oss) and information on \n\ masses and mixings. 

But what are these signatures? 

Very recently Duan, Fuller, Carlson, and Qian have pointed out \cite{Duan:2007sh} that,  due to the step in density, for \one\ \sne\ the  neutronization peak (a peak in the $\nue$ luminosity at about 10 ms post bounce) would not disappear regardless of the  value of $\t13$.  In contrast, for $\t13$ just below the current bound disappearance due to $\nue \rightarrow \numu,\nutau$ conversion is expected for \fe\ \sne\ \cite{Mikheev:1986if,Dighe:1999bi,Lunardini:2003eh}.  
Duan et al. also pointed out how the conversion pattern is more complex for a \one\ \sn, because \n-\n\ scattering influences the resonant conversion driven by matter. 
The initial study in \cite{Duan:2007sh}, as well as the analytical elaboration in \cite{Dasgupta:2008cd},  were restricted to the \n\ flux at the neutronization peak and to relatively large $\t13$. Beyond these, the subject of \n\ conversion inside a \one\ \sn\ remains to be explored.

Here we give a in depth  -- even though far from comprehensive -- study of \n\ \oss\ in \one\ \sne.  We present the first discussion of shock wave effects for this type of \sn, using a  numerical calculation of the density profile and of the neutrino fluxes, as they evolve over several hundreds of milliseconds.  

We describe a variety of conversion effects that could appear at different times depending on the value of $\t13$ and on whether
the detector is shielded by the Earth when the \n\ burst reaches it. 
We also address the question of what features could be observable and how.
A first  goal is to outline what combination of signatures one
should look for to test  models of \one\ \sne; another purpose is to  
describe the phenomena that should be taken into account when
analyzing \n\ data from a \one\ \sn.

The paper opens with generalities on \one\ \sne\ and details on our numerical model (sec.\ref{starmodel}). In sec. \ref{oscill} we give a discussion of  conversion effects, followed by our  results in sec. \ref{results}. A discussion on the implications of our findings closes the paper in sec. \ref{disc}.


\section{Oxygen-neon-magnesium-core \sne: the model }
\label{starmodel}

\begin{figure}[htbp]
  \centering
\includegraphics[width=0.48\textwidth]{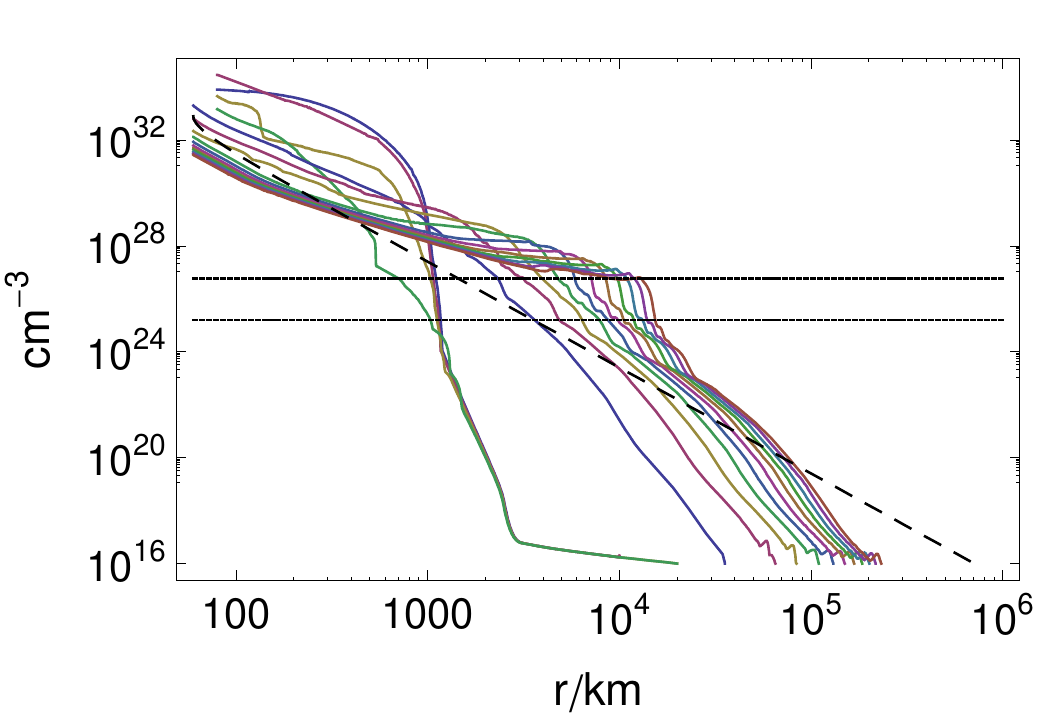}
\caption{Snapshots of the electron density profile at 
$t=0,\,50,\,100,....,\,700\,$ms (lines from top to bottom at their
left end, except for the inverted curves for $t = 0$
and $t = 50\,$ms). The positions of the supernova shock for $t \geq 200\,$ms 
coincide essentially with the lower right footpoints of the profiles. 
For $t=300\,$ms we also plot the effective number density of neutrinos 
(dashed curve, see Eq.~(\ref{effnu})), which is responsible for the effects of 
neutrino-neutrino forward scattering. The two horizontal lines represent the 
densities corresponding to the two MSW resonances for a neutrino
of 20$\,$MeV energy. }
\label{profiles_multi}
\end{figure}

The interior structure of stars in the range $\sim$8--10$\,M_\odot$ ($M_\odot=1.99 \times 10^{30}$ kg is the mass of the Sun)
is distinctively different from that
of more massive supernova progenitors: their ONeMg core is
surrounded by a thin carbon and oxygen shell and an even thinner
helium layer. The density gradient in this surface region of the
core (at around 1000~Km) is extremely steep; the density can drop
by nearly seven orders of magnitude within only 300$\,$Km.
This is visible from Fig.~\ref{profiles_multi}, where the
line corresponding to time $t = 0$ displays the electron
number density profile 
of the innermost region of the 8.8$\,M_\odot$
progenitor star used as initial model for the core-collapse 
and supernova simulations this work is based on (baryon 
densities in the core and its surface layer can be obtained 
by multiplication by a factor of 2). The ONeMg core
of this star was evolved to the onset of collapse by 
Nomoto \cite{nom84,nom87} and was more recently extended outside of the
thin He shell by a hydrostatic hydrogen envelope \cite{nomotopriv}.

The updated progenitor model was followed through collapse and explosion
with the neutrino-hydrodynamics code \textsc{Vertex} \cite{Rampp:2002bq,Buras:2005rp}. Details of
the employed input physics and information about the supernova dynamics
can be found in the works by Kitaura et al. \cite{Kitaura:2005bt} and Janka et al. \cite{Janka:2007yu,Janka:2007di}.
Figure~\ref{profiles_multi} provides an overview of the evolution of
the electron density ($n_e$) profile in steps of 50$\,$ms until 700$\,$ms 
after the start of a spherically symmetric (1D) simulation. The core bounce
happens at $t = 53.6\,$ms and the explosion sets in at about 130$\,$ms 
(to be recognized from positive velocities developing in the post-shock 
layer). 

For $t=100\,$ms the step in density induced by the shock is visible at 
$r \sim 130\,$Km. The shock front advances outward rapidly, reaching
$r \sim 500\,$Km at $t\lesssim 150\,$ms and meeting the base of the He shell  
(at $r \sim 1100\,$Km and $n_e \sim 4\times 10^{26}\,$cm$^{-3}$) 
less than 10$\,$ms later.  At this point, the shock accelerates even more
and travels essentially with the speed of light until it is gradually
decelerated again by running into the flatter density profile of the 
hydrogen envelope and sweeping up matter from the progenitor star.
At the end of the computed evolution, the shock has reached a radius 
of more than 2$\times 10^5\,$Km (Fig.~\ref{profiles_multi}).

As a consequence of the differential collapse of the core, starting with the
innermost regions and only gradually encompassing layers farther away from
the center, the density step is significantly flatter at the time when the shock  hits the ONeMg-core surface and passes through the base of the He shell.
For the first roughly
100$\,$ms after the shock passage through the step, the density profile
therefore appears essentially smooth. The ejecta in the region of the
previous density step then expand with nearly uniform and constant velocity 
($\sim$2--3$\times 10^4\,$cm$\,$s$^{-1}$) and the width of this layer 
(1000--2000$\,$km) is preserved. At later times, however, the profile appears
to steepen again because of the dramatically increasing radial scale and
the resulting relative decrease of the ratio of 
width to radius of the shell. The
relics of the original density step of the pre-collapse core can 
therefore still be recognized as an outward moving step-like structure 
on the density profile of the shock-accelerated and expanding ejecta 
of the supernova explosion.

\begin{figure}[htbp]
  \centering
 \includegraphics[width=0.4\textwidth]{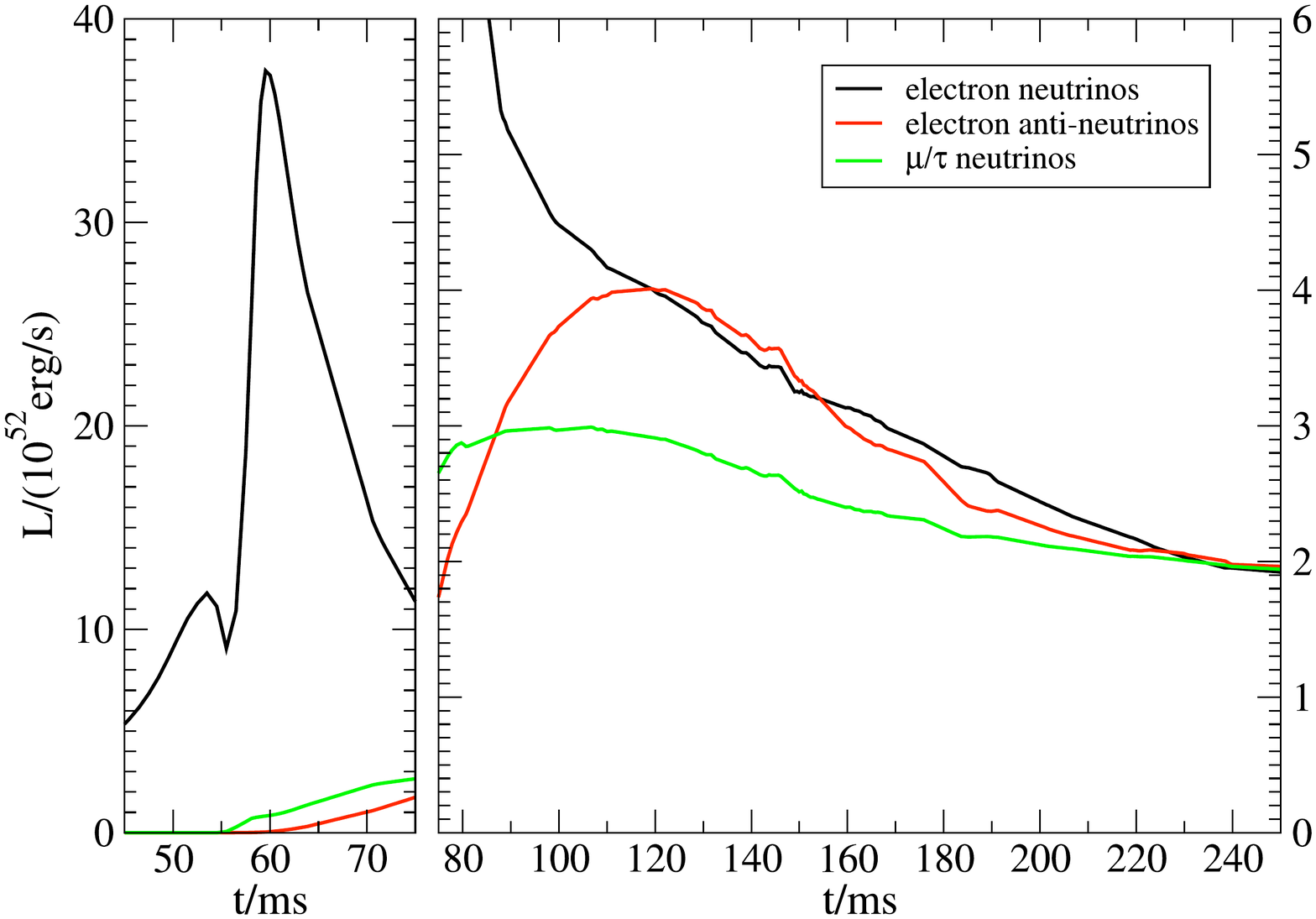}
 \includegraphics[width=0.4\textwidth]{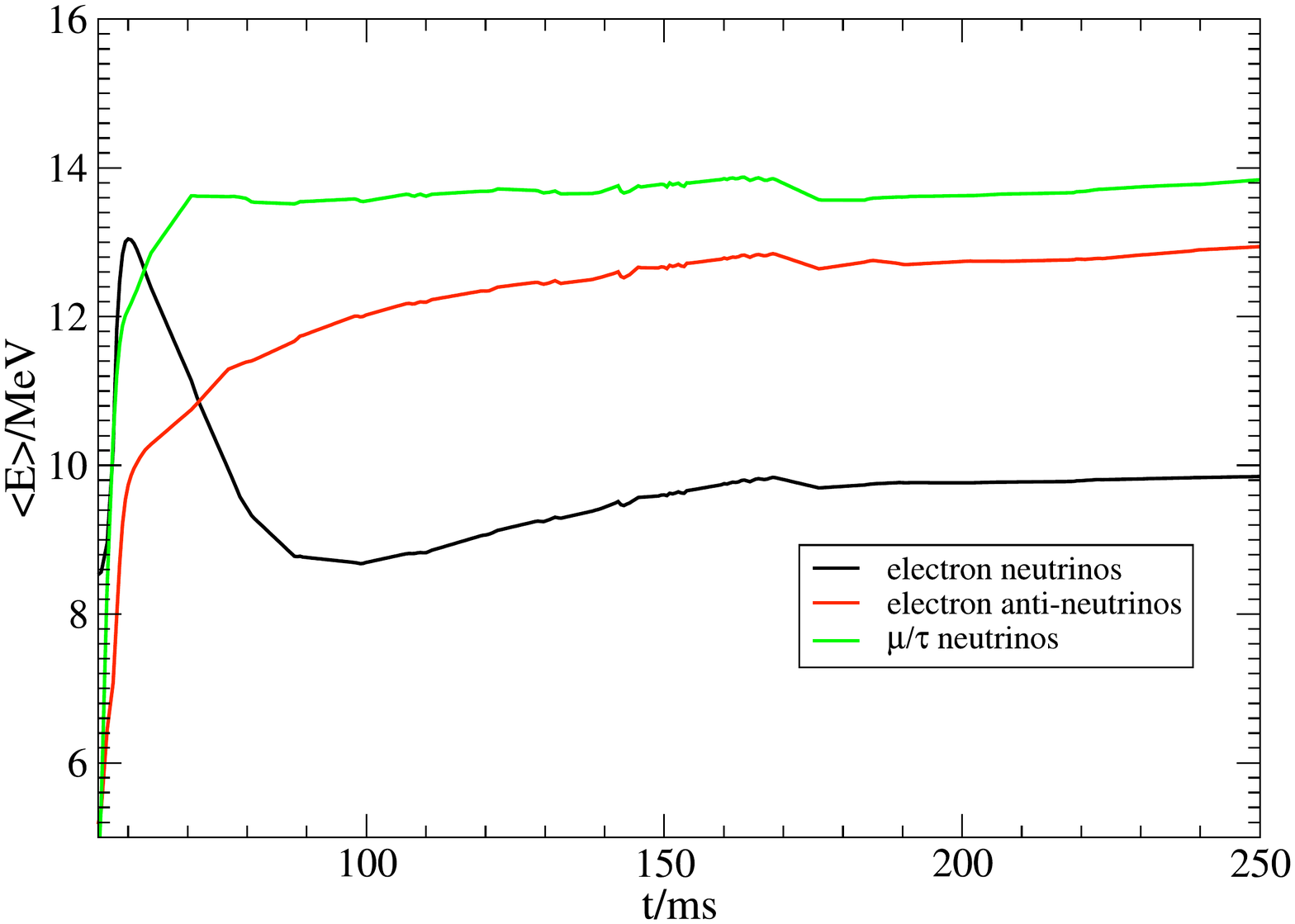}
\caption{Neutrino luminosities (top) and average energies (bottom)
before oscillations. The $\nux$ ($\nu_x=\numu,\nutau,\barnumu,\barnutau$) 
luminosity is per species, so the total luminosity in the non-electron 
flavors is four times the one plotted here. The average energy is 
defined as the ratio of energy flux to number flux.}
\label{lum_mean}
\end{figure}

The time evolution of the neutrino luminosities 
and of the mean spectral energies
(measured at a radius of $r=400\,$Km for an observer in the laboratory
frame), without including flavor conversion, is shown in Fig.~\ref{lum_mean}.
The displayed data are from a two-dimensional (2D) simulation (assuming 
axisymmetry around the polar grid axis) of an ONeMg-core supernova,
which was followed until roughly 200$\,$ms after core bounce. The
explosion dynamics of the 1D and 2D models is very similar and the 
density profiles as shown in Fig.~\ref{profiles_multi} basically also apply
for the 2D model after averaging over latitudes (thus wiping out 
inhomogeneities associated with convective overturn in the 
neutrino-heating layer around the neutron star). In contrast to 1D results,
the neutrino luminosities and mean energies from the 2D model include the
modifications caused to the neutrino emission by convective energy 
transport inside the nascent neutron star. This plays a role
later than a few ten milliseconds after core bounce and leads to 
slightly (about 10\%) enhanced electron neutrino and in particular 
muon and tau neutrino luminosities, a slightly reduced electron 
antineutrino luminosity, and somewhat lowered ($\sim$0.5--1$\,$MeV)
mean energies of all radiated neutrinos \cite{Buras:2005tb}.

Figure~\ref{lum_mean} shows that the luminosities, flavor composition,
and mean spectral energies of the neutrinos reveal the well known
behavior: the peak in $\nue$ associated with the neutronization burst,
the slower rise of the $\barnue$ and $\nux$
($\nu_x=\numu,\nutau,\barnumu,\barnutau$) luminosities, and the
hierarchy of the $\nue$, $\barnue$ and $\nux$ average energies. The
main difference compared to Fe-core supernovae is the absence of an
extended phase of post-bounce accretion of the forming neutron star
and therefore the lack of the corresponding accretion plateau in the
neutrino luminosities after the $\nu_e$ shock-breakout burst.

The 2D simulation was only carried to about 200$\,$ms after bounce.
For our analysis of neutrino oscillation effects at later times
until the end of the 1D simulation at $\sim 650\,$ms post bounce,
we will take the neutrino luminosities and mean energies to be roughly
constant. This is a reasonably good assumption because of the mentioned 
absence of accretion and the only slow, quasi-steady evolution of the 
settling neutron star. Moreover, our discussion will mostly concentrate on
matter-of-principle effects and the details of the neutrino emission 
properties will not be relevant.

The fluxes in Fig. \ref{lum_mean} will be used as initial conditions
for the calculation of \n\ oscillations, making the assumption that 
the neutrino luminosities and mean energies are constant outside of the
neutrinosphere.

In our calculations, we model the \n\ energy spectra following Keil, Raffelt and Janka 
\cite{Keil:2002in}:
\be
 \frac{{d} N}{{d} E}\simeq \frac{(1+\alpha)^{1+\alpha}L}
  {\Gamma (1+\alpha){E_{0}}^2}
  \left(\frac{E}{{E_{0}}}\right)^{\alpha}
  e^{-(1+\alpha)E/{E_{0}}},
  \label{nuspec}\ee
 where $E_0$ is the average energy and $L$ the luminosity. We take $\alpha=2.3$, for which the spectrum is close to a Fermi-Dirac distribution.  The power law in Eq. (\ref{nuspec}) is adequate in the fact that it reproduces well the numerical results of sophisticated \n\ transport (though by choosing a constant
value for $\alpha$ we neglect the detailed evolution of the spectral shape 
provided by the simulation results) \cite{Keil:2002in}.  

\section{Flavor conversion in the star and in the Earth: generalities}
\label{oscill}

Neutrino conversion in supernovae is particularly
complex due to the interplay of several effects.  
Here we review them briefly, in the measure needed to highlight the phenomena that are distinctive of a \one\ \sn.  We refer to the literature for more complete reviews (e.g., \cite{Dighe:1999bi,Kotake:2005zn,Fogli:2007bk}). 

The  flavor evolution of \ns\ inside the star and in the Earth is described by a Hamiltonian that contains three terms. The first is the kinetic term, which
depends on three mixing angles $\theta_{ij}$, a CP-violating phase $\delta_{CP}$, and  two independent
mass squared splittings, $\Delta m^2_{ji}$, where the indices
$i,j=1,2,3$ refer to the three mass eigenstates $\nu_1,\nu_2,\nu_3$.
The three characteristic oscillation frequencies associated with the
kinetic term are: 
\be 
\omega_{ji}=\frac{\Delta m^2_{ji}}{2E}~,
\label{freq_vac}
\ee
with $E$
the neutrino energy.  We use  the standard  parametrization
of the mixing matrix,  with the values $\sin^2 \theta_{12}= 0.32$, $\Delta m^2_{21}=8 \cdot 10^{-5}~{\rm eV^2} $,  $|\Delta
m^2_{31}|=3 \cdot 10^{-3}~{\rm eV^2} $,
 and $\s213 \leq 2 \cdot 10^{-2}$ (see e.g., the summary in \cite{GonzalezGarcia:2007ib} and references therein).     Notice that $\t13$ has not been
measured yet \cite{Apollonio:1999ae,Boehm:2001ik}, and that $\Delta m^2_{31}$ is known only in absolute
value. The case of positive (negative) $\Delta m^2_{31}$ is called
\emph{ normal} (\emph{inverted}) mass hierarchy.   The phase $\delta_{CP}$ can be ignored here, as it does not influence the relevant probabilities for equal $\numu$ and $\nutau$ fluxes  \cite{Akhmedov:2002zj,Balantekin:2007es}. 

The second and third  terms relevant for \n\ conversion are due to neutrino forward scattering (refraction) on electrons and on neutrinos. Their characteristic frequencies are 
\be
\omega_{e}=\sqrt{2} G_F n_e, \hskip 1truecm \omega_{\nu}=\sqrt{2} G_F n^{eff}_\nu~,
\label{freq_matt}
\ee
 where $n_e$
is  the number density of electrons along the trajectory of the \ns, and  
 $n^{eff}_\nu$ is the effective number density of neutrinos. It is defined as (see e.g., \cite{Fogli:2007bk}):
\be
n^{eff}_\nu = N_\nu 
\frac{1}{2}
\left[1- \sqrt{1- \left(\frac{R}{r}\right)^2}\,\right]^2~,
\label{effnu}
\ee
where $r$ is the radial coordinate, $R$ is the 
 radius of the neutrinosphere, and $N_\nu$ is the total number density of neutrinos and antineutrinos of all flavors at the neutrinosphere: $  N_\nu=\sum_{\alpha=e,\mu,\tau}(N_\alpha + \bar N_{\alpha})$.
The  self-interaction term from \n-\n\ scattering generates non-linear effects that only recently have been studied in detail in the context of supernovae \cite{Duan:2005cp,Duan:2006an,Hannestad:2006nj,Duan:2007mv,Raffelt:2007cb,Duan:2007fw,Duan:2007bt,Fogli:2007bk,Raffelt:2007xt,Duan:2007sh,Dasgupta:2007ws,Duan:2008za,Dasgupta:2008cd}. 

The character of the \n\ conversion is roughly determined by the
relative size of the different frequencies $\omega_{ji}$,  $\omega_e$, and  $\omega_\nu$.  Each of the two
refraction terms is relevant if its frequency is comparable or larger
than the kinetic (vacuum) ones.  The more terms contribute simultaneously, the
higher is the complexity of the conversion pattern.  Fig. \ref{profiles_multi}
allows to compare the strengths of the vacuum and refraction terms.
Together with  the electron density profile at different times, the figure shows  the values of $n_e$ for which the conditions $\omega_{21}=\omega_e$ and $\omega_{31}=\omega_e$ are realized for $E=20$ MeV. For small mixing angles (as it is the case for $\theta_{13}$), these well approximate the condition for the  MSW resonance \cite{Wolfenstein:1977ue,Mikheev:1986gs,Mikheev:1986wj,Mikheev:1986if}.  We also plot  the effective neutrino number density, $n^{eff}_\nu(r)$ for $t=300$ ms and $R=60$ Km.   This quantity is smaller (larger) at later (earlier) times, as follows from fig. \ref{lum_mean}. 

From fig. \ref{profiles_multi} we identify two different scenarios: the
stage before the shock reaches the He shell, $t\lta 150$ ms, and the later
times, $t\gta 150-200$ ms. These phases will be called \emph{pre-shock } and \emph{ post-shock } respectively from here on.  In what follows we discuss them for the propagation of neutrinos with the normal mass hierarchy, which is the most transparent and sufficient to illustrate the main effects.

\subsection{The late stage: post-shock}

Let us first consider the late, post-shock phase, for $t \gta 300$ ms.
For this, in the most internal part of the star ($r\simeq 100$ Km)
both the \n-electron and the \n-\n\ scattering terms exceed the vacuum
one and are comparable to each other.  In general, they both affect
oscillations substantially.  As the neutrinos propagate out to larger
radii, the influence self-interaction is strong over a
typical distance of $10^2 - 10^3$ Km.  At $r \simeq 3000$ Km, the
\n-\n\ term, $\omega_\nu$, falls below the \vac\ frequency
$\omega_{21}$, and thus becomes negligible, marking the end of
self-interaction effects.  Since the electron number density decreases
more slowly with the radius, at this distance the \n-electron term is
still large and dominates over the \vac\ one by at least one order of
magnitude.  Only at radius $r\gta 4-5 \cdot 10^3$ Km we have
$\omega_{31} \sim \omega_e$, which drives the inner MSW resonance.
Another resonance, driven by $\omega_{21}\sim \omega_e$, follows at
even larger radius, $r \gta 8 \cdot 10^3$ Km, beyond which all matter
effects are negligible.  The separation between the distance of
decoupling of the self-interaction effects and the position of the MSW
resonances, as well as the separation between the two resonances
themselves, increases with time.  Thanks to this separation, the phase
of collective effects, driven by \n-\n\ scattering, and the MSW
resonances are decoupled.

A scenario with these characteristics  has been studied, for a \fe\ \sn,  by Fogli et al. \cite{Fogli:2007bk}. They  found that,  beyond the region where self-interaction effects end, the \n\ spectra are unchanged for the normal mass hierarchy.   
Such conclusion has little dependence on $\t13$ and on the specific electron density profile, as long as the condition $\omega_e\gg \omega_{31}$ is satisfied over the distance of effectiveness \n-\n\ scattering  \cite{Fogli:2007bk} (see also \cite{Hannestad:2006nj,Duan:2007mv} for further discussions).  It is supported by analytical arguments based on the minimization of potential energy of a spin in a magnetic field (e.g.,  \cite{Hannestad:2006nj,Raffelt:2007cb,Fogli:2007bk,Raffelt:2007xt}).  Only with the violation of the matter-dominance condition (which is not our case) a swap of the $\nue$ and $\numu/\nutau$ spectra might occur below a critical energy $E_C\lta 10$ MeV \cite{Duan:2007bt}. Even in the implausible case that a swap with $E_C\lta 10$ MeV  occurs in our scenario, 
such swap would depend only indirectly on the electron density, with no strong sensitivity to the density structure of a \one\ \sn.   

For all these reasons, we consider it adequate here to consider
unchanged neutrino spectra immediately after (in radial distance) the end
 of self interaction effects, and focus of the physics of the MSW-driven
transformation, which, thanks to its direct and resonant dependence on
$n_e$, is a much more sensitive probe of the matter distribution
inside the star.  As will be seen, the most interesting effects of the
MSW resonances are at high energy ($E \gta 30$ MeV), and thus are not
influenced by a possible swap at low energy.

After the MSW-driven conversion, the flux of electron neutrinos at Earth, $F_e$, can be written in terms of the fluxes without oscillations, $F^0_e$ and $F^0_x$ ($x=\mu,\tau$), as:
\be
F_e=p F^0_e + (1-p) F^0_x~,
\label{fluxearth}
\ee
where $p$ denotes the  $\nue$ survival probability.  This probability is given by \cite{Dighe:1999bi}:
\be p \simeq P_H \left[\cos^2 \theta_{12} P_L + \sin^2 \theta_{12}
(1-P_L) \right]~.
\label{p}
\ee
Here $P_H$ ($P_L$) is the probability  of  transition between the eigenstates $\nu_3$  and $\nu_2$ ($\nu_2$ and $\nu_1$) of the Hamiltonian in the higher density (lower density) MSW resonance.
$P_H$ depends on $\t13$ and on the derivative of the electron number density, calculated at the point $r_{res}$ where $\omega_{31}=\omega_e$  (see e.g., \cite{Fogli:2001pm}):
\begin{eqnarray}
&&P_H = \frac{e^{\chi \cos^2 \theta_{13}} -1}{e^{\chi} -1}~,
 \nonumber \\
&&\chi \equiv - 2 \pi \frac{\Delta m^2_{31}}{2 E} \left[\frac{1}{n_e
(r)} \frac{dn_e(r)}{dr}\right]^{-1}_{r=r_{res}}~.
\label{phform}
\end{eqnarray}
$P_L$ obeys the same expression with the replacements $\omega_{31}\rightarrow \omega_{21}$, $\theta_{13} \rightarrow \theta_{12}$ and $\Delta m^2_{31}\rightarrow \Delta m^2_{21}$.
We can see that the survival probability varies between $p=0$  and $p=\cos^2 \theta_{12}$ (numerically, between $p=0$ and  $p=0.68$), for $P_H$ and $P_L$ varying in the allowed range from 0 (\emph{adiabatic} propagation) to 1 (completely \emph{non-adiabatic} propagation).  The derivative of the density profile in Eq. (\ref{phform}) encodes the sensitivity of the MSW effect to the matter distribution: the amount of flavor conversion depends on how rapidly the electron density changes along the neutrino trajectory and on the size of the mixing angle.  Steeper profiles and smaller mixings correspond to stronger violation of adiabaticity. 

The post-shock phase can be divided in an \emph{early} part, $t \lta 650$ ms, and a \emph{late} one, $t\gta 650$ ms. The late part begins when, accelerated by the shock, the He shell has traveled outwards and expanded to lower density, to a degree that the density step associated with it does not extend to the resonance density anymore, and thus becomes irrelevant for \n\ conversion.  In coincidence with this transition to a shallower density profile, we expect  the MSW conversion to become more adiabatic.  Such early transition in the direction from less to more adiabatic conversion is unique of a \one\ \sn.

\subsection{The early stage: pre-shock}
\label{preshock}

Let us now discuss the more complicated pre-shock phase. Here, all the three
frequencies  are comparable at $r \simeq 10^3$ Km.  Therefore
the physics of the MSW resonances can not be decoupled from that of
\n-\n\ scattering as in the post-shock phase. Moreover, the two
resonances are spatially closer, to the point that the factorization of
their effects may not be valid.  

In ref. 
\cite{Duan:2007sh} the $\nue$ survival probability for this case was calculated numerically, for $\t13=0.1$. It appears that, for the normal hierarchy, the probability can be described by transitions between the eigenstates
of the Hamiltonian, in a way that resembles the simpler case of two spatially
separated resonances without effects of \n-\n\ refraction. 
An important feature is the presence of sharp swaps in the energy spectrum, instead of the smoother transitions between adiabatic and non-adiabatic regimes that are expected in the pure MSW case as the neutrino energy increases.  

In the limit of slowly varying neutrino density, the probability found in \cite{Duan:2007sh} has an analytical interpretation \cite{Dasgupta:2008cd}: it is the result of collective MSW transitions -- where the collective behavior is due to \n-\n\ scattering -- followed, at lower density, by spectral swaps analogous to those already observed  for other density profiles \cite{Duan:2006an,Duan:2007bt}. 
The  $\nue$ survival probability 
can be described with the same expression, Eq.
(\ref{p}), using effective values for  $P_H$ and $P_L$
 \footnote{For simplicity of language, throughout the paper $P_H$ and $P_L$ will be referred to as 
``jumping probabilities'', 
and the expression ``adiabatic''  for the neutrino propagation will be used in connection to them.  We stress that, while this terminology is accurate for the MSW resonances in the post-shock case, in the pre-shock phase it is simply an effective description, and does not correspond to the physics of the \n\ propagation, due to collective effects.   }.  In first approximation, these behave as step functions, whose critical energies are determined by conservation laws \cite{Duan:2008za,Dasgupta:2008cd}.
Specifically, the numerical result in \cite{Duan:2007sh} is reproduced by:
\beq
&&P_H=\begin{cases}
1 & \text{for $E>12$ MeV}\\
0 & 
\text{otherwise}
\end{cases}\nonumber \\
&&P_L=\begin{cases}
1 & \text{for $E>15$ MeV}\\
0 & 
\text{otherwise}
\end{cases}
\eeq
%
%
so that we have $p\simeq 0$ below 12 MeV and $p \simeq 0.68$ at $E \gta  15$ MeV, with $p\simeq 0.32$ as intermediate value between the two.   

While the dependence of the results on $\t13$ was not shown in  \cite{Duan:2007sh}, it was checked by the same authors \cite{duanpriv} that the behavior $p= \cos^2 \theta_{12} \simeq 0.68$ at high energy (above a critical energy $E_C \lta 15$ MeV) is a generic feature even for smaller $\t13$, reflecting the fact that the produced $\nue$'s always emerge the star almost completely in the lightest mass state, $\nu_1$.   This can also be understood considering that in the limit $\t13 \rightarrow 0$, the third mass state is a pure mixture of $\numu$ and $\nutau$, and therefore it decouples from the evolution of $\nue$.  This implies that the $\nue$ survival probability should depend only on the pair of states $\nu_2$ and $\nu_1$.  These have a $\nue$ content ranging from $\sin^2 \theta_{12}=0.32$ to $\cos^2 \theta_{12}=0.68$. A $\nu_2 \rightarrow \nu_1$ swap would produce a change in the survival probability from the lower to the higher of these two values.

Here we use the numerical result from ref. \cite{Duan:2007sh} for the survival probability.  For $\t13=0.1$, it applies well to our case because all the parameters match:  our pre-shock density profile  matches the one in  \cite{Duan:2007sh} and the \n\ spectrum at the neutronization peak ($t \simeq 60$ ms) has a very similar average energy (12-13 MeV against the 11 MeV in  \cite{Duan:2007sh})  \footnote{A major, but trivial, difference  with respect to the plots of ref. \cite{Duan:2007sh} is due to the fact that we use the probabilities at Earth, averaged over oscillations, while the plots in \cite{Duan:2007sh} refer to a radial distance of 5000 Km.}.  For smaller $\t13$, we consider the same result to be still applicable to the high energy range, in the light of the reasoning above.   As it will appear shortly, the high energy range is where we expect the most robust signatures of conversion, namely the Earth regeneration.

Notice that we do not address the \n\ conversion at  the instants
of time that immediately precede and follow the shock passage through
the density step, $t \sim 150-300$ ms.  This is  because in this interval the conversion pattern is highly complicated: the matter near the
resonance density undergoes a quick acceleration to relativistic
velocity  with the passage of the shock, thus making relativistic corrections necessary \cite{Nunokawa:1997dp}.  Instead, for the same density, velocities are lower than 10\% of the speed of light at later times.  In addition to relativistic corrections, one would need to include  all the oscillation terms (kinetic, electron scattering and self-interaction) simultaneously, as in the pre-shock phase.
The combination of
 these effects would require a dedicated study that is postponed for the time being.

\subsection{Oscillations in the Earth}

In addition to conversion in the star, \ns\ from a \sn\ undergo oscillations inside the Earth.  These oscillations are observable by a detector that is shielded by the Earth when the \n\ burst reaches it. They  are driven by \n-electron scattering. The  phase of oscillation  depends on the solar parameters, $\theta_{12}$ and $\Delta m^2_{21}$, while the amplitude reflects the pattern of conversion in the star. 
The difference between the $\nue$ flux in a detector with and without Earth shielding has the expression (valid for normal hierarchy) \cite{Dighe:1999bi,Lunardini:2001pb}: 
\be
F^D_e - F_e = (F^0_e - F^0_x) P_H(1- 2 P_L) f_{reg}~.
\label{diffearth}
\ee
Here $f_{reg}$ is the \emph{regeneration factor}. Up to terms proportional to $\sin^2 \theta_{13}$, $f_{reg}$ is given by:
\be
f_{reg}=P^{\oplus}(\nu_2 \rightarrow \nue)-\sin^2\theta_{12}~,
\label{regf}
\ee
where  $P^{\oplus}(\nu_2 \rightarrow \nue)$ is the probability that a state entering the Earth as $\nu_2$ is detected as $\nu_e$ in the detector.  In the absence of Earth shielding, $P^{\oplus}(\nu_2 \rightarrow \nue)=\sin^2\theta_{12}$ and  $f_{reg}=0$.
The regeneration factor  is an oscillatory function, and $f_{reg} \gta 0$
with good approximation at all energies and zenith angles.  Notice
that Eq. (\ref{diffearth}) exhibits an elegant factorization of the
several steps of the \n\ propagation from production to detection: the first factor describes the originally produced fluxes, the second the
higher density resonance, the third the low density resonance and the
fourth the effect of the Earth.  This factorization makes the Earth effect particularly transparent, and the oscillatory character makes it an unambiguous signature of \n\ flavor conversion, that can not be confused with astrophysical effects.

Here we calculate the \oss\ in the Earth using Eq. (\ref{diffearth}) with  $P_L$ and $P_H$ as described above for the two regimes (pre and post shock), and with  $f_{reg}$ from the accurate numerical calculation in \cite{Lunardini:2001pb}. 


\section{Conversion of neutrinos from an ONeMg-core \sn: results}
\label{results}

\subsection{Probabilities}
\label{prob_results}

From the previous sections, one can see two oscillation effects that
are distinctive of a \one\ \sn.  One of them is the influence on the
adiabaticity of conversion of the sharp gradient in density that marks
the oxygen-helium transition in the star.  One expects strong
adabaticity breaking even for relatively large $\t13$, in contrast to
\fe\ \sne.  The second effect is the time variation of the oscillation
probabilities within the first hundreds of milliseconds of the burst,
associated with the variation of the density profile with the shock
passage. In \fe\ \sne\ shock effects appear at much later times. Below
we illustrate these effects in detail and compare them to the results for \fe\ \sne\ from the literature (e.g.,
\cite{Dighe:1999bi,Kotake:2005zn,Fogli:2007bk}).
In our calculations we have
used the results of Duan et al. \cite{Duan:2007sh} for the pre-shock
phase, and the density profiles in fig. \ref{profiles_multi}, together
with the analytics in sec. \ref{oscill} for the post-shock times.

\begin{figure}[htbp]
  \centering
 \includegraphics[width=0.4\textwidth]{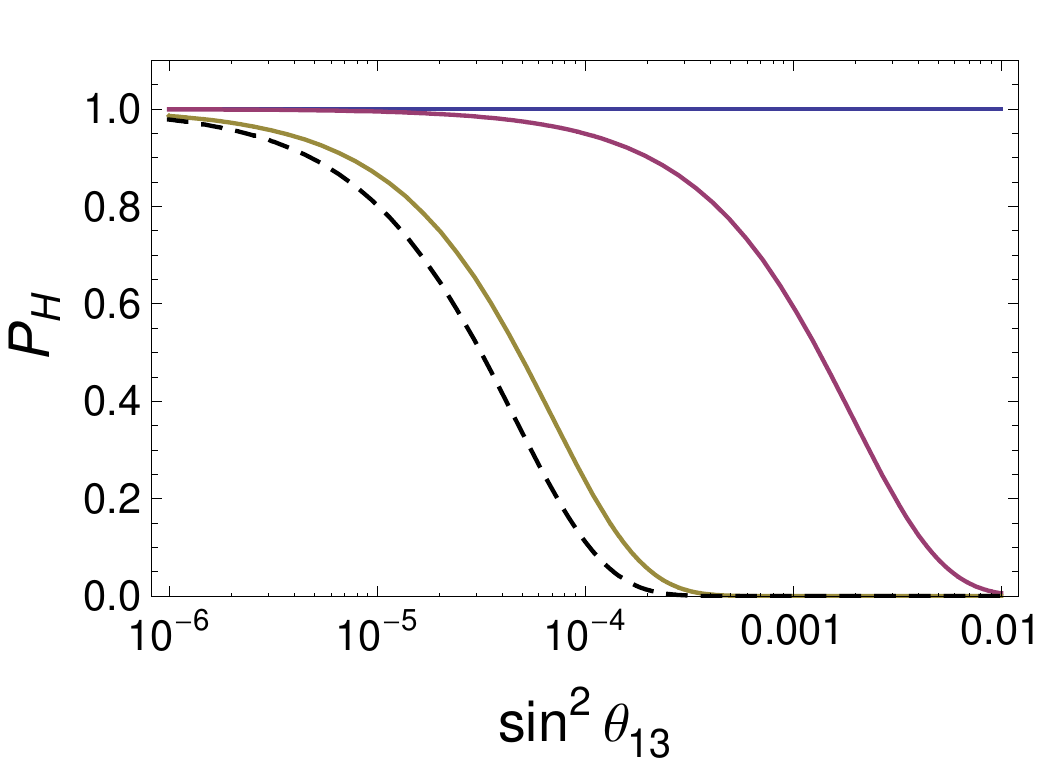}
\caption{The jumping  probability $P_H$ for $t=60,450,700$ ms (solid curves, from upper to lower) as a function of  $\sin^2 \theta_{13}$ for energy $E=20$ MeV. The dashed line shows the same probability for a Fe \sn\ with the parameters in ref. \cite{Lunardini:2003eh}.  } 
\label{phfig}
\end{figure}

Let us begin with the jumping probability $P_H$.  
In fig. \ref{phfig} $P_H$ is plotted  as a function of $\s213$ for $E=20$ MeV and for three representative times, one in the pre-shock phase ($t=60$ ms \footnote{The choice of 60 ms is motivated by its marking the peak of the neutronization flux.  The density profile was not calculated for this precise instant of time, and so the profile at $t=50$ ms has been used as a very reasonable approximation.}), one in the early post-shock  phase ($t=450$ ms) and one in the late   post-shock stage ($t=700$ ms). The same function for a \fe\ \sn\ (at $t\lta 3$ s, from ref. \cite{Lunardini:2003eh}) is also shown for comparison.

We see essentially three regimes in $\t13$: 

\begin{enumerate}

\item  \emph{large}: $\sin^2 \theta_{13}\gta 3 \cdot  10^{-3} $.  The character of the conversion changes fast (within $\sim 200-300$ ms or so), turning from  non-adiabatic in the pre-shock phase, to completely adiabatic afterward. 

\item  \emph{intermediate}: $\sin^2 \theta_{13}\simeq  3 \cdot 10^{-5} - 3 \cdot 10^{-3} $ The character of conversion changes more slowly: it is
  completely non-adiabatic in the pre-shock phase, and after $\sim 300$ ms it is still at least partially non-adiabatic. Only in the late post-shock phase ($t\gta 700$ ms)  it may become completely adiabatic if $\sin^2 \theta_{13}\sim 3 \cdot  10^{-4} - 3 \cdot 10^{-3} $.  In this range of mixing the  three time intervals (pre-shock, post-shock early and post-shock late) can be distinguished. 
  
\item \emph{small}: $\sin^2 \theta_{13}\lta 3 \cdot  10^{-5}$.  The conversion
  remains completely non-adiabatic at all times, with only a minor
  change in the transition from the early to the late post-shock
  phases.

\end{enumerate}

Notice that only in the late post-shock stage the dependence of $P_H$ on $\t13$ resembles the one of a \fe\ \sn, for which there is no time variation until much later times.

The jumping probability in the low density resonance, $P_L$, is found
to be zero at all post-shock times, and is set to 1 before the shock for $E\gta 12$ MeV,
as discussed in sec. \ref{oscill}.  For a \fe\ \sn\ $P_L=0$ at all times.

\begin{figure}[htbp]
  \centering
\includegraphics[width=0.4\textwidth]{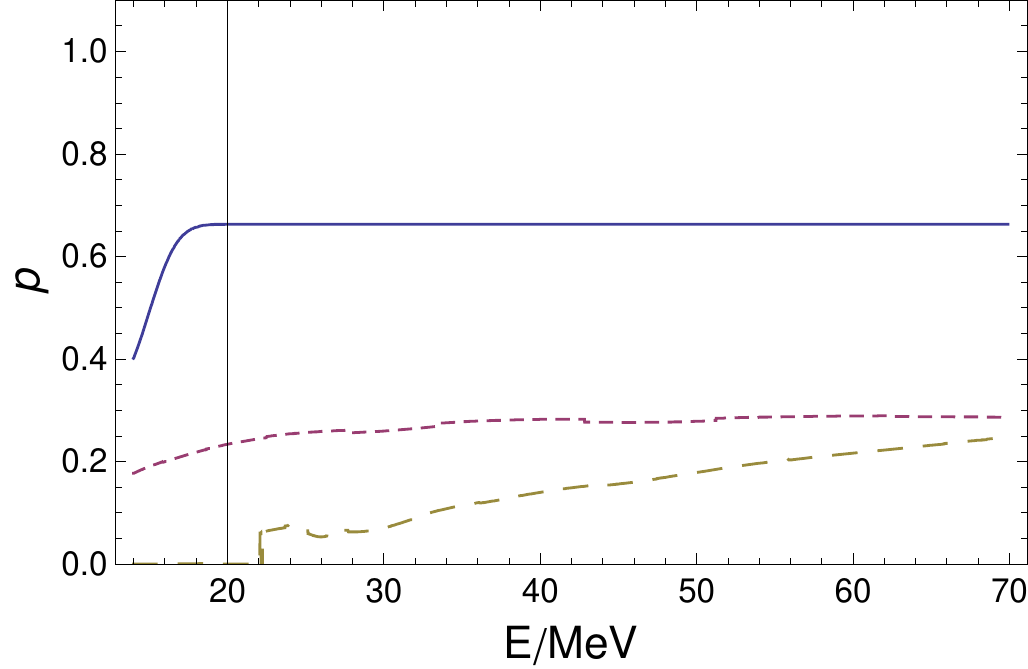}
\caption{The electron neutrino survival probability for $t=60,450,700$ ms (solid, short dashed, and long dashed, respectively) and $\sin^2 \theta_{13}=  6 \cdot10^{-4}$. $p=0$ for a \fe\ \sn\ over the same time interval.} 
\label{psurv_multi}
\end{figure}
The behavior of $P_L$, combined with the results in fig. \ref{phfig} for $P_H$,
explains what we see in fig.  \ref{psurv_multi}, showing the energy
dependence of the survival probability $p$  for $\sin^2
\theta_{13}=6 \cdot  10^{-4}$ and no Earth shielding.  In the transition from the pre- to the
post-shock phases the probability decreases from $\sim 0.68$ to $\sim
0.32$ (asymptotic values at high energy) mostly as a consequence of
the change in $P_L$.  At later times, $p$ goes to zero, beginning with
the lower energy part of the neutrino spectrum.  This appears in fig. \ref{psurv_multi}: for $t=700$ ms below $E \sim 22$ MeV we have
$p=0$, because for these energies the high density resonance is
realized in the shallower part of the density profile, more internal
with respect to the base of the He shell (regrouped after the shock passage). Instead, at higher energy the
resonance is realized at lower density, on the density step.  The
step is moving outwards and becoming less dense, so the transition
 to $p \simeq 0$ moves to larger energy.

For small $\t13$, $P_H\simeq 1$ at all times, and so the only change in $p$ is due to the transition of $P_L$:  the asymptotic value of $p$ goes  from $\sim 0.68$ to $\sim 0.32$ and retains this value at all later times. 

For large $\t13$, the probability $p$ changes from $p\simeq 68$ to zero immediately after the shock passage, when $P_H$ goes to zero. 

\begin{figure}[htbp]
  \centering
 \includegraphics[width=0.45\textwidth]{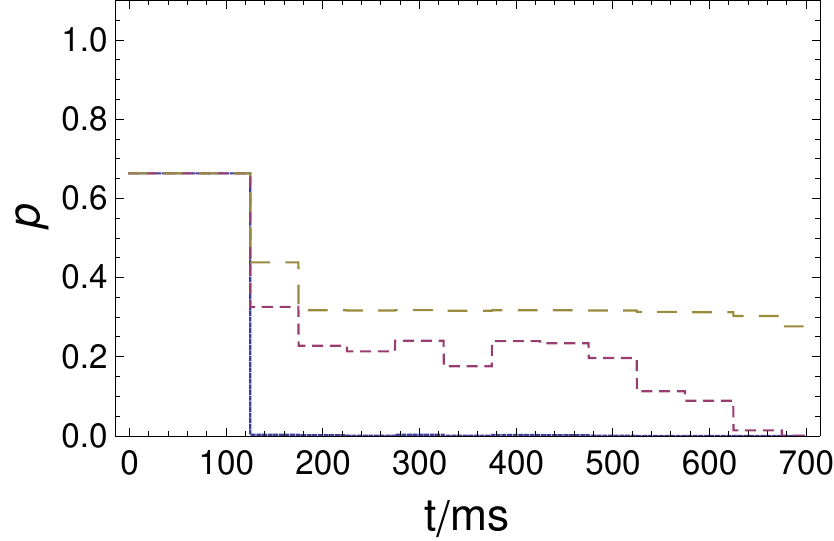}
  \caption{The $\nue$ survival probability as a function of time for $E=20$ MeV and   $\sin^2\theta_{13}=0.01,6 \cdot 10^{-4}, 10^{-5}$ (solid, short dashed, and long dashed, respectively).  The step-like structure  reflects the fact that the oscillation effects were evaluated for density profiles in steps of 50 ms ($t=0,50,100,..$ ms, middle points of the steps, see profiles of fig. \ref{profiles_multi}).  For  $t \sim 150-300$ ms the figure is only schematic: it does not capture the full complexity of the conversion in accelerating matter when the shock passes through the density step (see sec. \ref{preshock}).
  } 
\label{psurvtime}
\end{figure}

The features seen in fig. \ref{psurv_multi} appear also in fig. \ref{psurvtime}, where the time dependence of $p$ is shown for fixed energy and different values of $\theta_{13}$.  The figure is a good quantitative description for the pre-shock and late post-shock phases ($t \gta 300$ ms), while for the intermediate times it has only a schematic character. This is because it does not include a number of effects that are relevant when the matter at the base of the He shell is strongly accelerated by the shock (see sec. \ref{preshock}). The figure evidences clearly the general decline of the survival probability with time, up to minor deviations from the monotonic trend due to the minute details of the density profile at the MSW resonance. It is also apparent that this decrease is faster for larger $\t13$, as already noticed above.

The conversion pattern found for a \one\ \sn\ is in contrast with that for a  \fe\ \sn\ over the same time interval: in that case, the survival probability has no time dependence and ranges from $p=\sin^2 \theta_{12}\simeq 0.32$ to $p=0$ depending on $\theta_{13}$ (fig. \ref{phfig}).

In the absence of Earth shielding, the observation of the oscillation effects could be challenging.  Indeed, the measurement of a probability would be complicated by uncertainties in the original \n\ fluxes, and the shock-induced modulations of the probabilities could be masked by the natural time evolution of the \n\ spectra and luminosities. 
The best signature to look at would probably be the  fate of the peak in the $\nue$ luminosity at $t\simeq 60$ ms: for large $\t13$ it  disappears for a \fe\ \sn, but it survives for an \one\ one \cite{Duan:2007sh}.   A conclusion on this would require knowing $\t13$, however.  Looking for the progressive vanishing of $p$ from lower to higher energy (fig. \ref{psurv_multi}) might also be promising. 

As it has already been pointed out for a \fe\ \sn\ \cite{Lunardini:2001pb,Dighe:2003jg,Dighe:2003vm}, the Earth shielding can be a great advantage, because 
oscillations in the Earth have an unambiguous signature: oscillatory
distortions in the observed \n\ energy spectrum.  The very presence of
the effect allows one to conclude on \n\ masses and mixing, even in
the presence of uncertainties on the original \n\ fluxes.  

\begin{figure}[htbp]
  \centering
 \includegraphics[width=0.4\textwidth]{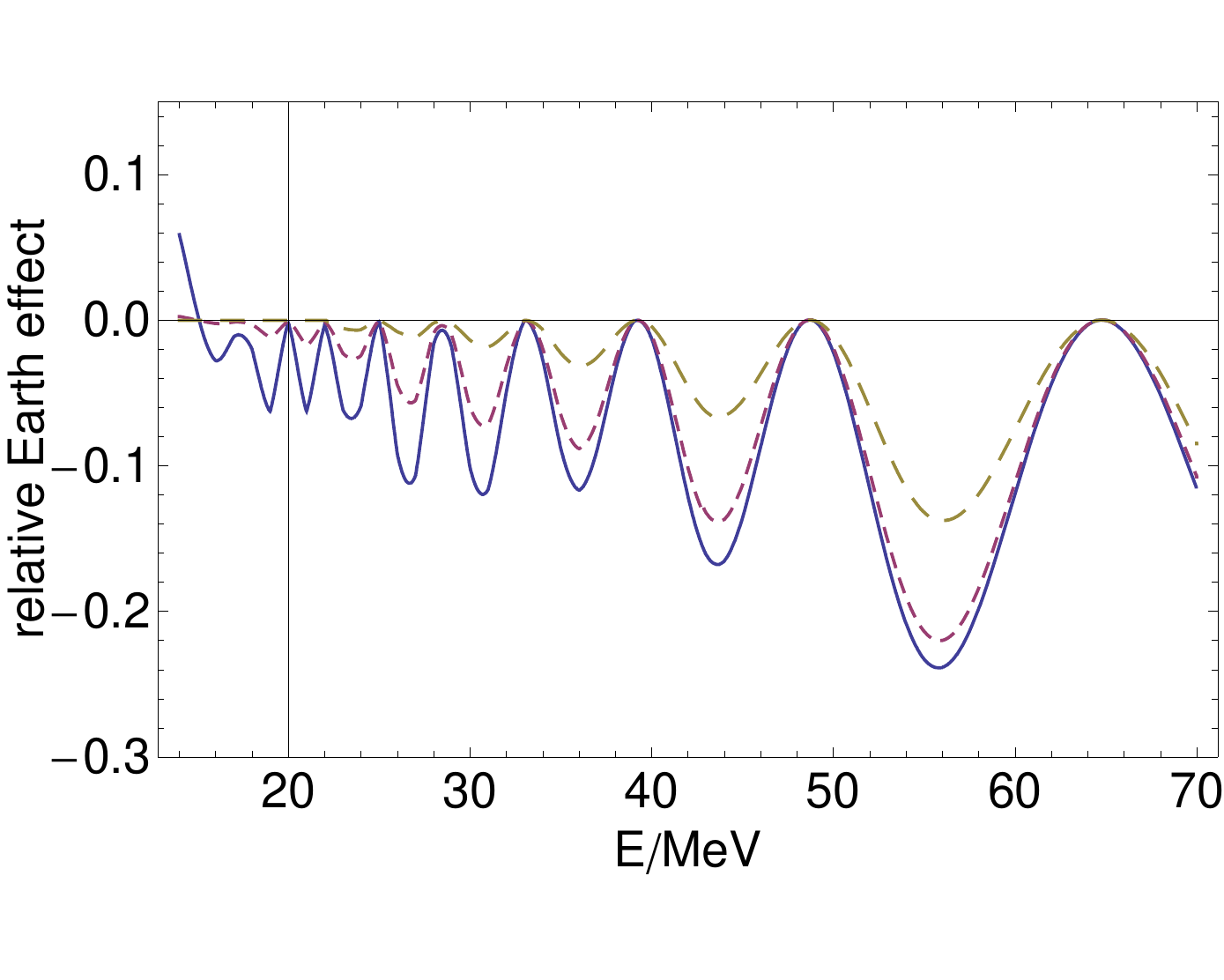}
\caption{The relative Earth matter effect in the neutrino channel, at 60 degrees nadir angle, for $t=60,450,700$ ms (solid-, short-dashed and long-dashed, respectively)  and $\sin^2 \theta_{13}= 6 \cdot 10^{-4}$.  
%
%
For a Fe \sn\ and the same value of $\theta_{13}$, the Earth effect is zero. } 
\label{relative_earth_0006}
\end{figure}
Fig. \ref{relative_earth_0006} shows the relative Earth effect, defined as $(F^D_e-F_e)/F_e$ (see Eq. (\ref{diffearth})), for nadir angle of $60$ degrees,  an intermediate value of $\t13$ ($\sin^2 \t13=6 \cdot 10^{-4}$) and the same instants of time as in the previous plots.  The amplitude of the effect reflects the time evolution of $P_H$ and $P_L$. In the pre-shock regime the amplitude is maximal, reaching $\sim 25\%$ effect at $E\simeq 56$ MeV. 
The effect is negative because, with $P_L\sim 1$, the term $(1-2 P_L)$ is negative and the flux contribution is positive, $F^0_e-F^0_x>0$, reflecting the fact that $\nue$ dominates the flux at this stage. 

In the early post-shock phase the effect is reduced in amplitude at
low energy, where the conversion is partially adiabatic in the high
density resonance.  Notice that, with respect to the pre-shock phase,
above $E \sim 15$ MeV there is a double change of
sign in Eq. (\ref{diffearth}): one is caused by $P_L$ changing from 1 to 0,
and the other due to the transition from $\nue$ dominated to
$\nux$-dominated (at high energy) flux, meaning a change of sign in the term $F^0_e-F^0_x$. 
 The net Earth effect is again 
negative.  Notice that the double change of sign is a unique feature of a \one\ \sn:  for any value of $\theta_{13}$ at $t=60$ ms the relative Earth effect would be positive  in a Fe \sn. 

We stress that the two sign flips may occur at slightly different
times, thus giving rise to an interesting sequence of sign changes in
the Earth effect.  It might also be possible to see the amplitude of
the oscillations become smaller and then increase again if the transition of $P_L$  from 1 to 0 is gradual and $P_L\simeq 1/2$ for a significant period of time and/or if the
quantity $F^0_e-F^0_x$ reaches a minimum in absolute value. 
To study if and how these time modulations happen is beyond the scope of the present paper: it would require information on the time evolution of the density profile with smaller time steps, and 
to use the time evolved profile for a  detailed numerical study of the neutrino conversion in a regime where all the three oscillation terms are relevant.

In the late post-shock phase the Earth effect is further reduced in
amplitude and ultimately goes to zero together with $P_H$, starting
with the lower energy part of the neutrino spectrum and including
higher energies as time passes. In Fig. \ref{relative_earth_0006} we see the same feature at $E =22$ MeV as in fig.  \ref{psurv_multi}, marking the transition to completely adiabatic conversion in the high-density resonance at late post-shock times. 

\begin{figure}[htbp]
  \centering
 \includegraphics[width=0.4\textwidth]{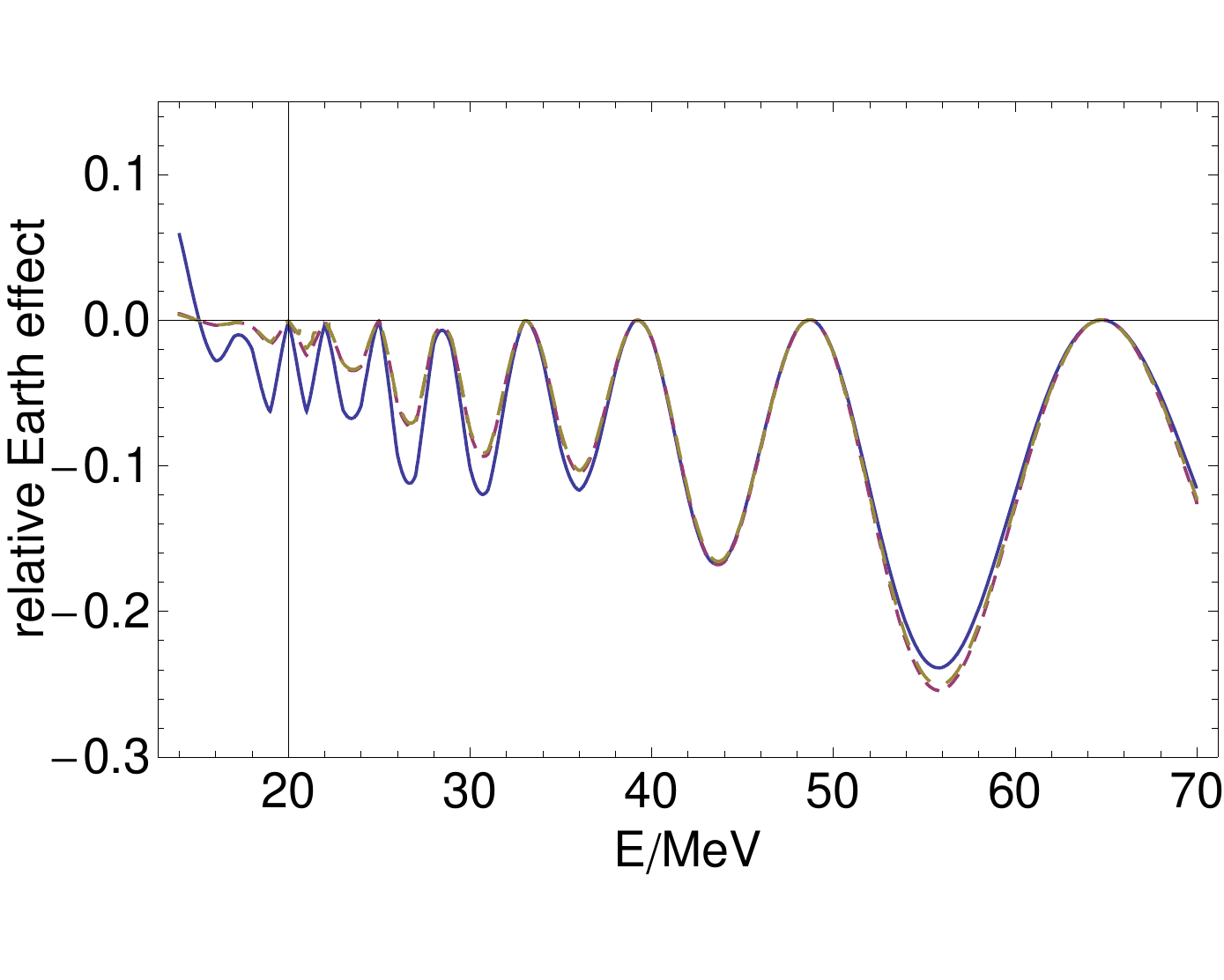}
\caption{Same as fig. \ref{relative_earth_0006} for $\sin^2 \theta_{13}=  10^{-5}$. The conversion is non-adiabatic at all times, and so the oscillations in the Earth never vanish.  They differ from the case of a \fe\ \sn\ in the sign at early times ($t \sim 60$ ms, around the neutronization peak): negative for \one\ \sne\ and positive for \fe\ ones.} 
\label{relative_earth_00001}
\end{figure}
If $\t13$ is small, the Earth effect is non-vanishing at all times
(fig. \ref{relative_earth_00001}), because the high-density resonance
remains completely non-adiabatic. 
The double sign flip discussed above happen in this
case as well, implying the same possibility for amplitude modulation
and/or sign changes as for the scenario with intermediate $\t13$. 
The major difference with respect to a \fe\ \sn\ is the different sign of the Earth effect at $t\sim 60$ ms, while at later times the two cases are similar.
\begin{figure}[htbp]
  \centering
 \includegraphics[width=0.4\textwidth]{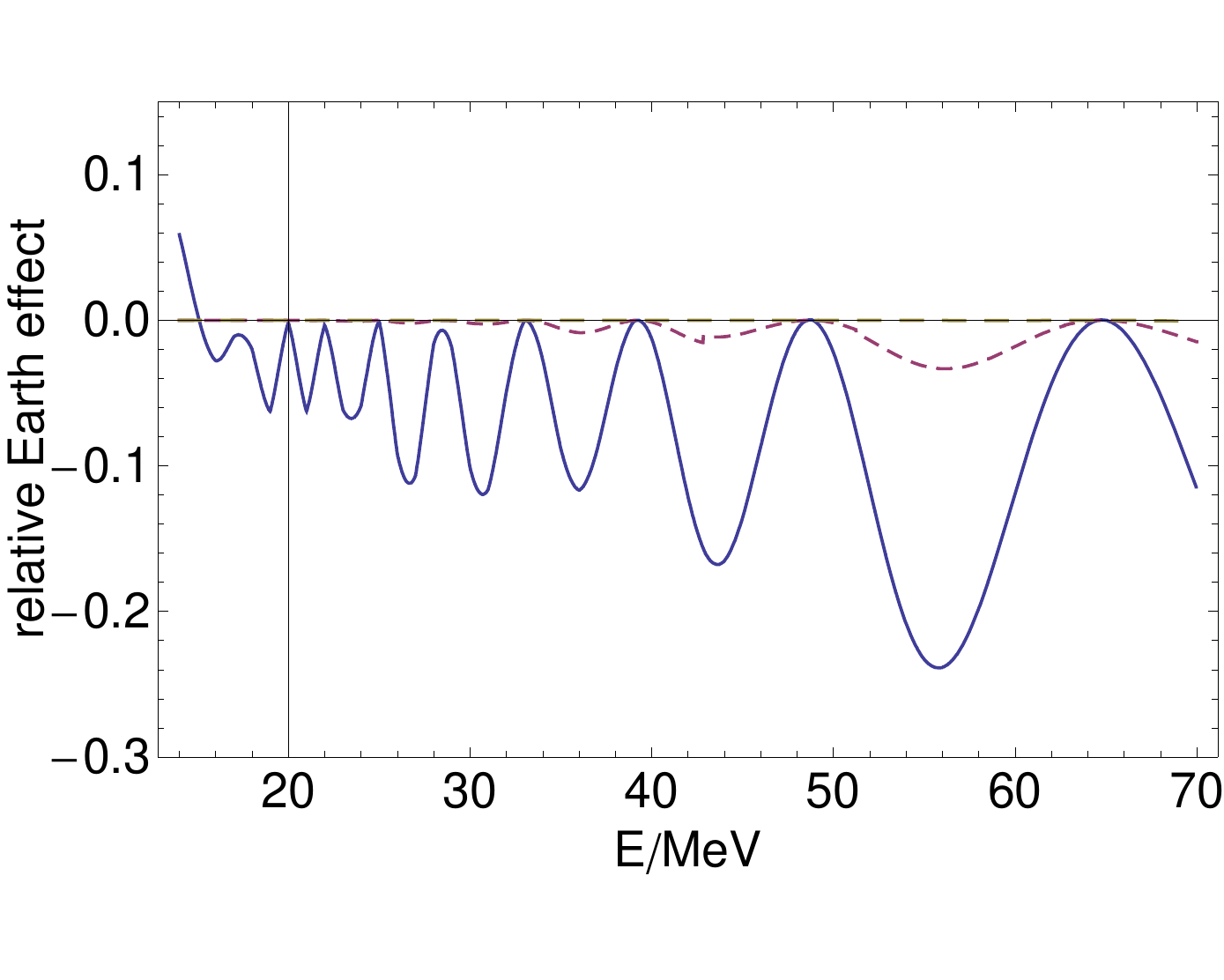}
\caption{Same as fig. \ref{relative_earth_0006} for $\sin^2 \theta_{13}=  10^{-2}$. After the shock reaches the He shell, the conversion is essentially adiabatic and the Earth effect disappears.  In a Fe \sn\ the effect would be zero at all times.} 
\label{relative_earth_01}
\end{figure}

For large $\t13$, oscillations in the Earth practically disappear with the shock
passage through the high-density resonance (fig.
\ref{relative_earth_01}), due to the adiabatic character of this
resonance in the post shock regime. This differs strongly from the
case of a \fe\ \sn, where for the same values of $\t13$ (and the same
normal mass hierarchy) the Earth effect is zero at all times.

The sudden or progressive disappearance of the Earth oscillations is
unique of a \one\ \sn.  For a \fe\ \sn\ the effect is either
constantly non-zero or it appears at late times ($t\gta 5$ s) due to  the shock wave propagation through the resonance layer
\cite{Lunardini:2003eh}.

We find similar results for other directions of propagation inside the Earth. The Earth effect is stronger  for deeper  trajectories inside the Earth, reaching about 40\% size at $\sim 70$ MeV for neutrinos moving along the diameter of the Earth. 

\subsection{Spectra}
\label{spectra_results}

Finally, we find it useful to show  the energy spectrum of the electron \n\  flux in a detector for a \one\ \sn. It is given in the left column of fig. \ref{onespectra}, for different times and different values of $\theta_{13}$.  The right column of the figure illustrates the same spectra, but with the oscillation effects calculated for a \fe\ \sn\ (from \cite{Lunardini:2003eh}) \footnote{ We emphasize that the results in both columns of fig. \ref{onespectra} were calculated using the same original fluxes shown in fig. \ref{lum_mean}, the only difference being in the flavor conversion.  While not realistic (the original spectra are different for \one\ and \fe\ \sne), this is useful for the illustration of the distinctive oscillation effects of the two types of \sne}.

The figure shows two dramatic signatures of a \one\ \sn\ in the high luminosity $\nue$ flux from neutronization ($t \simeq 60$ ms), that were already pointed out in \cite{Duan:2007sh}:  the lack of suppression of the  flux, regardless of the value of $\theta_{13}$, and  the step in the spectrum at $E\sim 12 $ MeV, due to \n-\n\ scattering.   The exact position and shape  of the step in the energy spectrum would require a full detailed calculation that is not done here. However, using the results in \cite{Dasgupta:2008cd} we have checked that these quantities are rather independent of the details of the original neutrino spectra and of the matter density profile, and this assures the validity of fig. \ref{onespectra} for illustration.

A third, well visible, feature is the negative sign of the Earth effect in the neutronization peak for a \one\ \sn\,in contrast with the positive sign for a \fe\ one.  This  could be precious to distinguish the two progenitor types if $\theta_{13} $ is small or unknown: indeed for small $\t13$ the neutronization peak is unsuppressed for both types and the step in the spectrum for the \one\ \sn\ could be masked by the small statistics and poor energy resolution if it is near the threshold of the detector (e.g., a 7 MeV threshold for a water Cherenkov detector).  The Earth effect, instead, is largest at high energy, where the energy resolution is better.

The features discussed for a \one\ \sn\ in the post-shock phase, namely the progressive decrease of the survival probability and of the Earth effect for intermediate $\t13$, are present, but not visible at the scales used in fig. \ref{onespectra}.  They are  of the order of $\sim (5 -25)\%$ depending on the energy and of the specific value of $\t13$. 

Notice that, while $p$ can decrease with time by a factor of 2 or more (fig. \ref{psurv_multi}), the net effect on the \n\ spectrum is modest. This is due to the fact that $p$ is small in value:  the observed $\nue$ flux is dominated by the originally produced $\numu,\nutau$ fluxes, and therefore it is impacted only at the level of tens of per cent by even major changes in the surviving $\nue$ component.

\section{Discussion}
\label{disc}

\begin{figure}[htbp]
  \centering
 \includegraphics[width=0.5\textwidth]{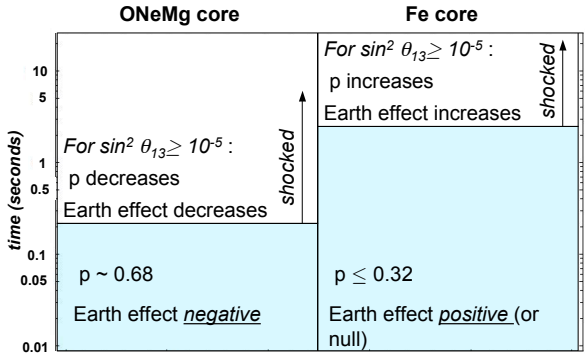}
  \caption{Summary of the  oscillation signatures of \one\  \sne\ compared to \fe\ \sne\ at different times.  They refer to the $\nue$ channel for the normal mass hierarchy; $p$ is the $\nue$ survival probability. The shaded areas represent the pre-shock phase, defined as the time interval before the shock front reaches the location of the MSW resonances, while the white regions (``shocked") indicate the later times.  
 } 
\label{summarytable}
\end{figure}

ONeMg-core collapses amount to 4\% to 20\% of all supernovae in the
local universe \cite{Poelarends:2007ip}. If one of them happens in our
galactic neighborhood, the observed conversion effects will be a
unique way to confirm the presence of the step in density at the base
of the He shell and the faster shock propagation relative to a \fe\
\sn.  The main oscillation signatures for both \sn\ types are summarized in fig. \ref{summarytable}. 
In more detail, the presence of the step characteristic of a \one\ \sn\  would be confirmed if: 

\begin{itemize}

\item  $\t13$ is known to be large and the neutronization peak does not disappear. Disappearance is predicted for the smooth profile of a \fe\ \sn, where the \n\ propagation is completely adiabatic. 

\item  $\t13$ is intermediate, small, or unknown, and data indicate a $\nue$ survival probability larger than $\sim  0.32$ in the first 100-150 ms of the burst. This indicates that at least part of the produced $\nue$'s exit the star in the state $\nu_1$, which can not happen for the smooth density profile of a  \fe\ \sn.

\item The Earth effect is negative in coincidence with the neutronization $\nue$ peak ($t\sim 60$ ms). This is another sign that $\nue$'s are converted into $\nu_1$'s.

\item  Shock effects -- regardless of when they happen -- evidence a change from less adiabatic to more adiabatic conversion as time passes.  Such a change manifests itself as decrease of the $\nue$ survival probability (increase of $\nue$-$\nux$ permutation) and decrease in the amplitude of the oscillations inside the Earth. In absence of the step, the shock passage would have the opposite effect \cite{Schirato:2002tg}. 

\end{itemize}

The timing of the  shock effects will test the scenario of the faster shock propagation unambiguously.  These effects  are characterized by features that move through the \n\ spectrum, from low to high energy, as time passes.  Such features can, in principle, provide a valuable measurement of the speed of the shock. 

The time-dependent features induced by the shock wave on the observed $\nue$ signal  will also give information on $\t13$.  In particular:

\begin{itemize}

\item  A large $\t13$ ($\s213 \gta 3 \cdot 10^{-3} $) will be singled out  if both the survival probability  $p$ and the amplitude of oscillations in the Earth drop quickly to zero in the post-shock phase (within $\sim 300$ ms).

\item  An intermediate $\t13$ ($\s213 \sim  3 \cdot 10^{-5} - 3 \cdot 10^{-3}$) is identified by the fact that both $p$ and the amplitude of oscillations in the Earth decrease more slowly, and may vanish only in the late post-shock phase, $t\gta 700$ ms.  This late transition will begin with the lower energy \ns\ and extend to higher energy as time passes.  
The probability $p$ changes in three distinct steps corresponding to  pre-shock, early and late post-shock. For $\s213 \sim  3 \cdot 10^{-4} - 3 \cdot 10^{-3}$  a similar three-steps behavior characterizes the Earth oscillations  at least in part of the energy spectrum (fig.  \ref{relative_earth_0006}).

\item A small $\t13$ ($\s213 \lta 3 \cdot 10^{-5}$) will appear from
  the fact that $p$ never vanishes, but only drops from $p\sim 0.68$
  to $p\sim 0.32$ after the shock reaches the base of the He
  shell.  The Earth oscillations also remain present at all times,
  either unchanged or with subtle time modulations.  These can be a
  temporary flip of sign and/or a temporary reduction or even
  disappearance of the amplitude. They are due to the change of $P_L$
  from 1 to 0, passing by the critical value of $P_L=1/2$, for which
  the Earth effect disappears.

\end{itemize}

In the absence of Earth shielding, one would need to measure the value of
the survival probability $p$ and look for its time variations.  Both
could be difficult because the \n\ fluxes in the different
flavors at the neutrinosphere are uncertain and evolve with time, a fact that can mask time features due to flavor conversion.
The non-disappearance of the neutronization peak and step-like features that move from low to high energy, like that in Fig. \ref{psurv_multi}, would probably be the easiest to distinguish.

If a detector is shielded by the Earth, instead, the chances to distinguish the early shock effects are much better, since the oscillatory distortions of the \n\ spectrum induced by the Earth can not be mimicked by any other phenomenon  \cite{Lunardini:2001pb,Dighe:2003jg}, and their phase is well known thanks to the relatively precise measurements of the parameters $\Delta m^2_{21}$ and $\theta_{12}$.

If a \sn\ is optically obscured, \n\ data will be the only source of
information on the nature of the progenitor. In such an event, searches for the oscillation signatures of an \one\ \sn\  would have an even higher importance.
Information from data at early times will be complemented and substantiated by the data at late times, $t \gta 5$ s, where shock effects are expected for a \fe\ \sn\ and $\t13$ in the intermediate and large range \cite{Takahashi:2002yj,Fogli:2003dw,Tomas:2004gr,Fogli:2004ff}.  For example, the late appearance of oscillations in the Earth would exclude an \one\ \sn.  
For small $\t13$ any shock wave signature disappears for a \fe\ \sn, making it 
more challenging  to distinguish between   \fe\ and \one\ types.  Discrimination is still possible from looking for  signs of non-zero $P_L$ in the pre-shock phase, which is typical of an \one\ \sn.  These are a large $\nue$ survival probability, $p\gta 0.32$, and the negative Earth effect around the neutronization peak.
 
In any event, the oscillations discussed here will be crucial for the
correct interpretation of data from an \one\ \sn, aimed at
reconstructing the original \n\ fluxes, which are so important to test
the theory of core collapse, reconstruct the mass of the neutron star
and -- indirectly -- its equation of state, and to discuss the conditions for
supernova nucleosynthesis.

Before concluding, a word of caution is necessary about the validity of our results.
They were obtained assuming that the conversion of \ns\ in the star
can be described in terms of transitions between the eigenstates of
the Hamiltonian -- which justifies Eq. (\ref{p}) --
with the transition probabilities as discussed in sec. \ref{oscill}.  While
all the literature available at this moment supports this
prescription, a confirmation from a detailed numerical study of the
problem still lacks.  
We expect that, after such study is done, our
results remain valid at least in the main message that a distinctive
pattern of \n\ conversion is associated with a \one\ \sn,
characterized by high survival probability of $\nue$ due to the step in density
at the base of the He shell and by early shock effects that drive the
conversion towards stronger flavor conversion.

\begin{widetext} 

\begin{figure}[htbp]
  \centering
 \includegraphics[height=0.3\textheight,width=0.33\textwidth]{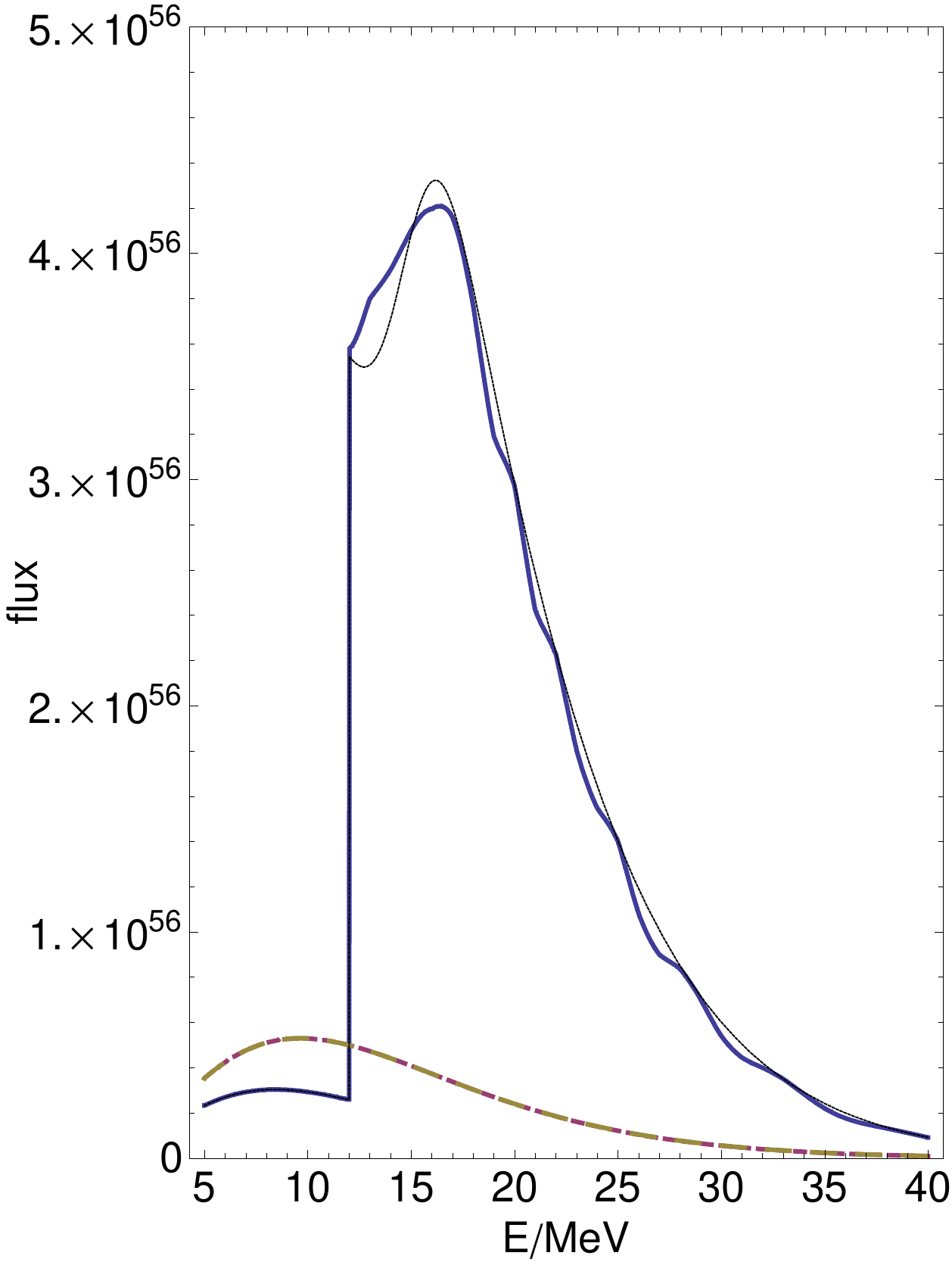}
\includegraphics[height=0.3\textheight,width=0.33\textwidth]{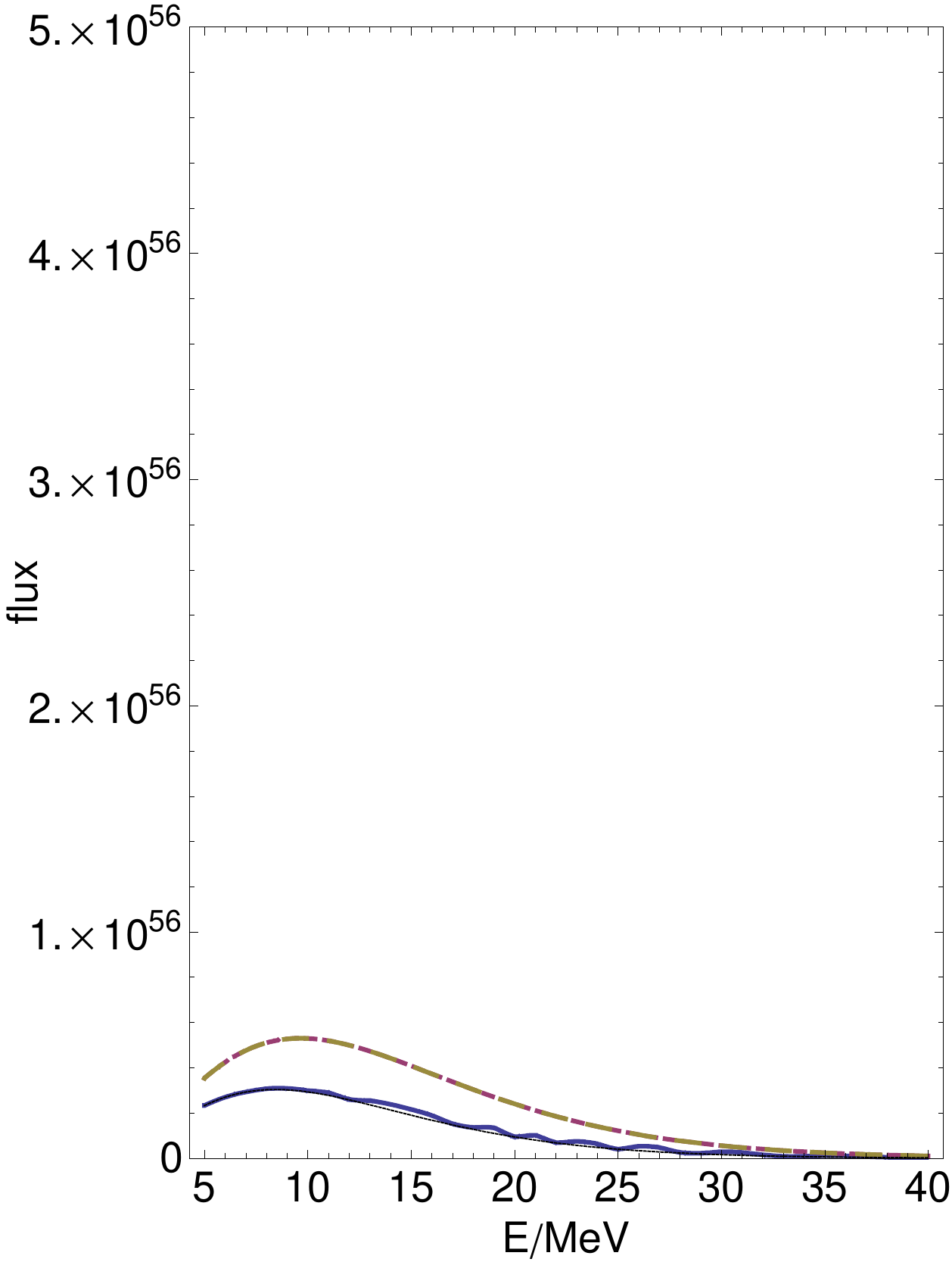}
 \includegraphics[height=0.3\textheight,width=0.33\textwidth]{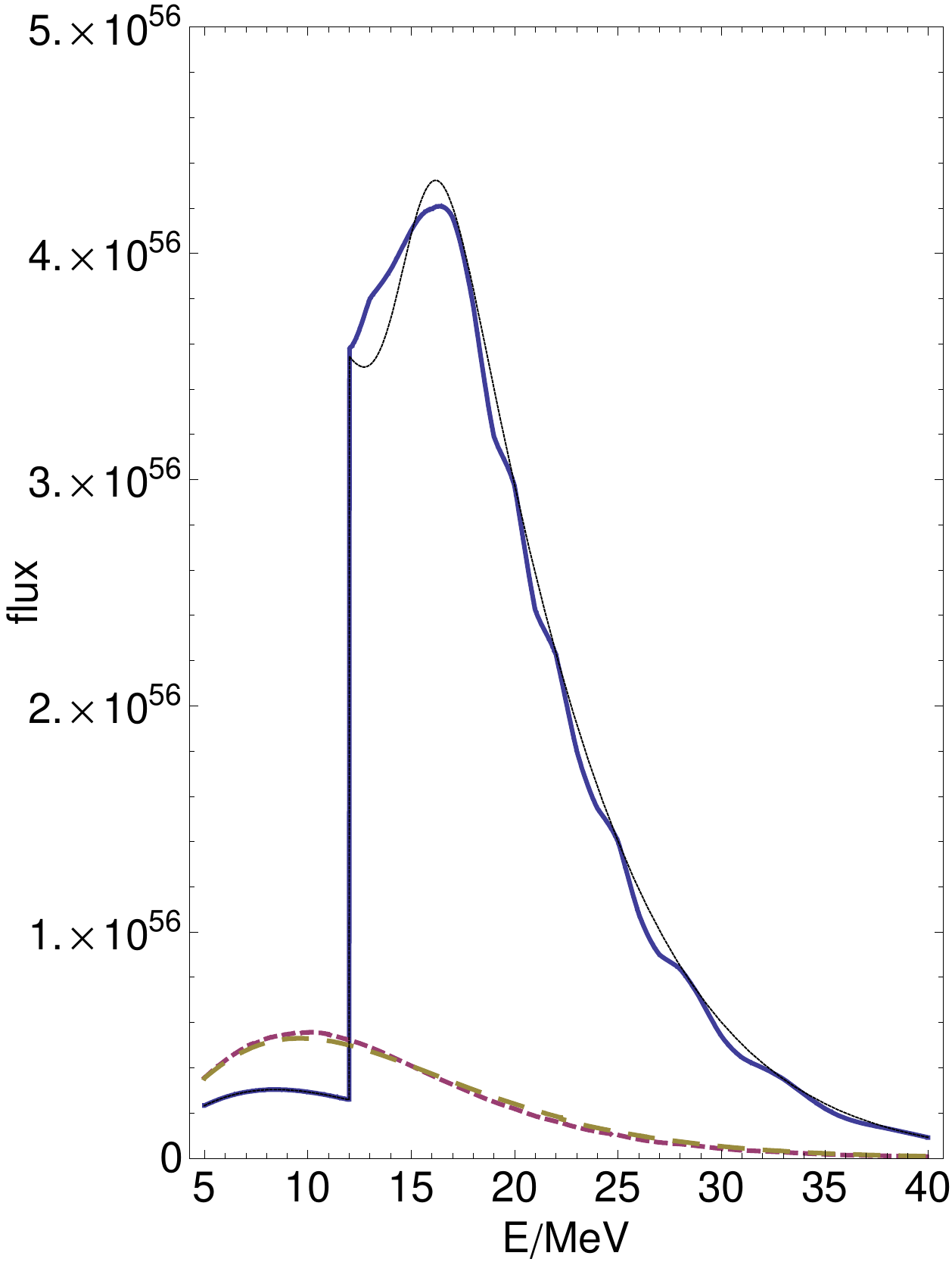}
\includegraphics[height=0.3\textheight,width=0.33\textwidth]{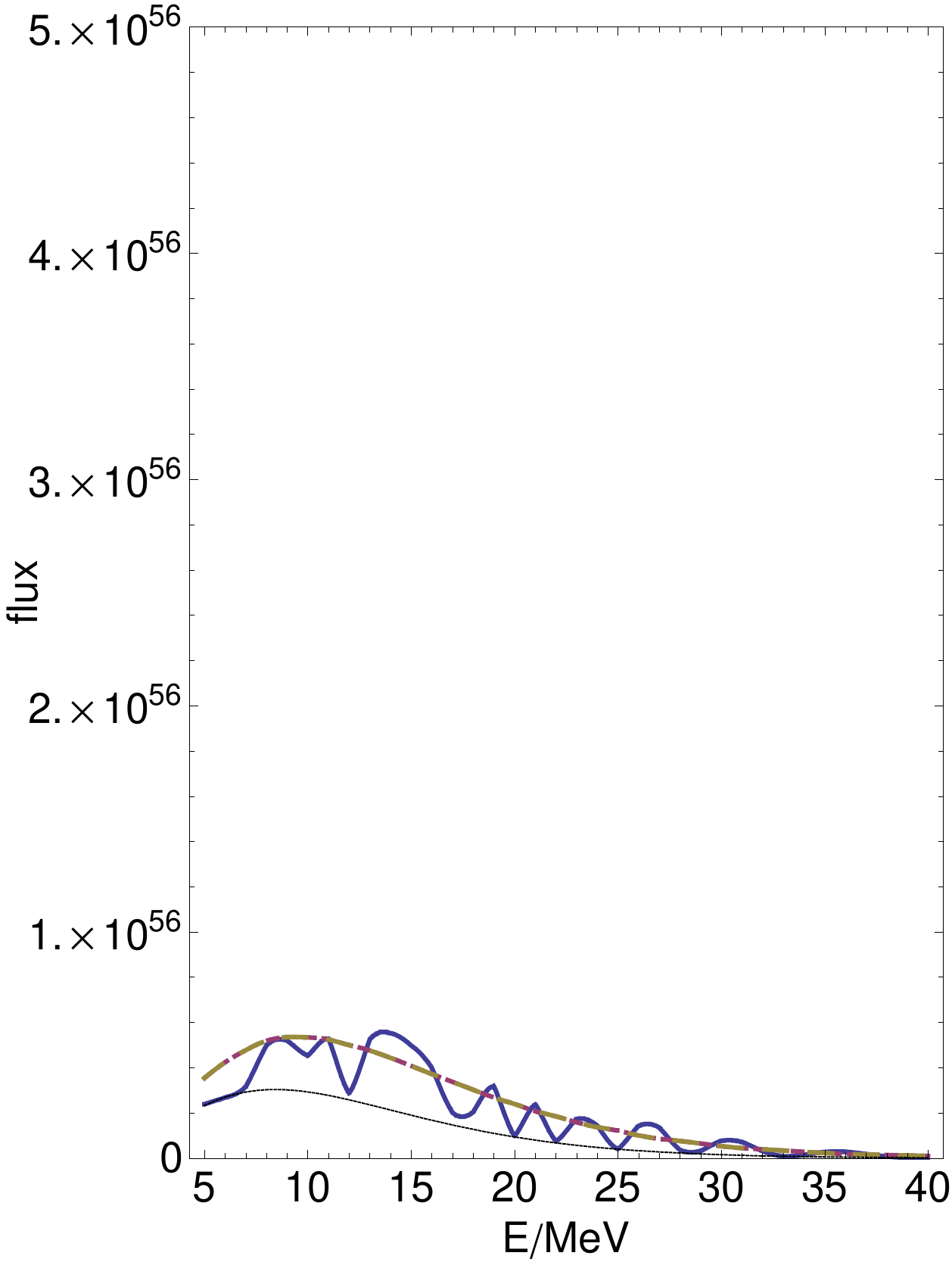}
  \includegraphics[height=0.3\textheight,width=0.33\textwidth]{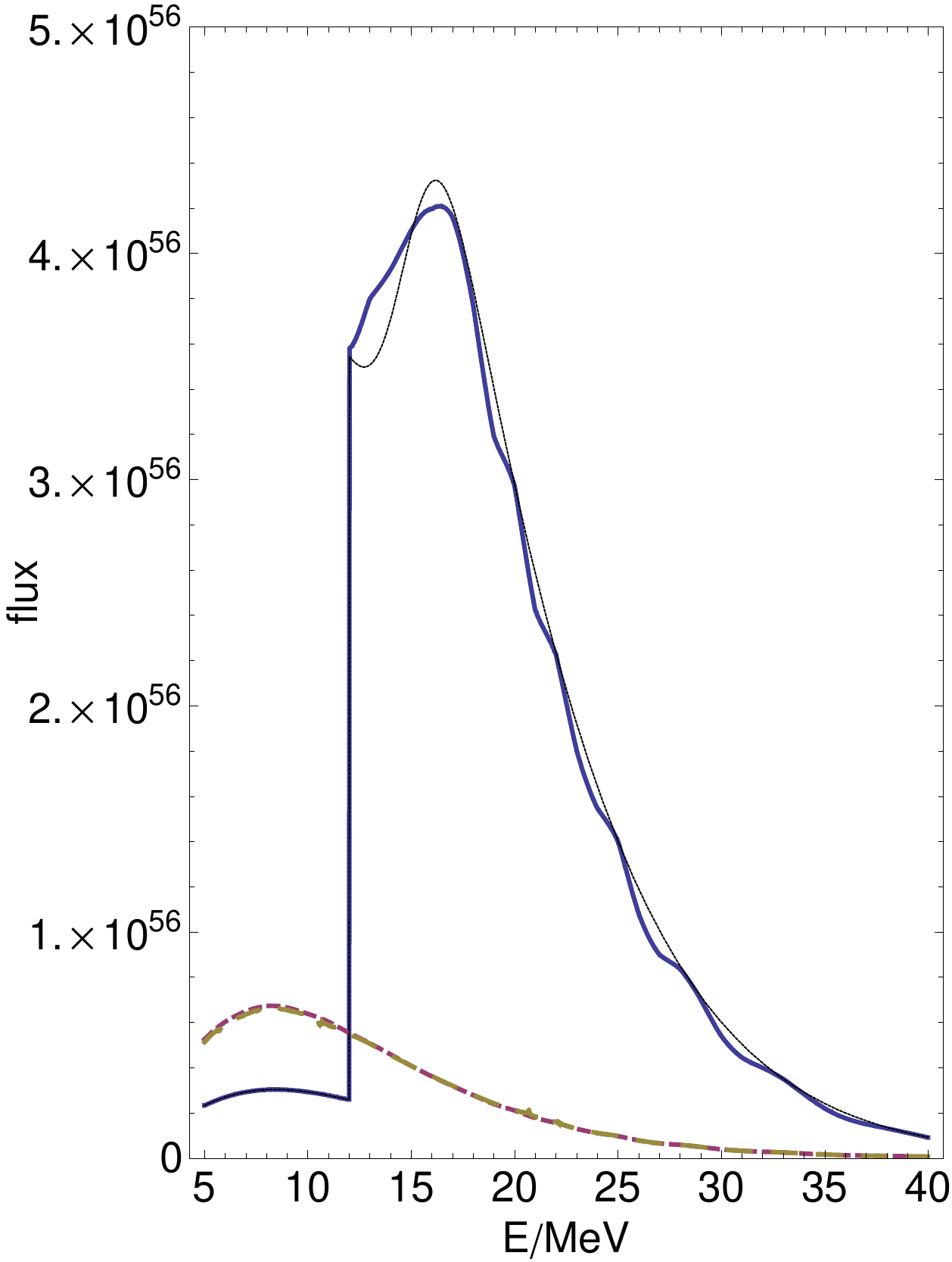}
  \includegraphics[height=0.3\textheight,width=0.33\textwidth]{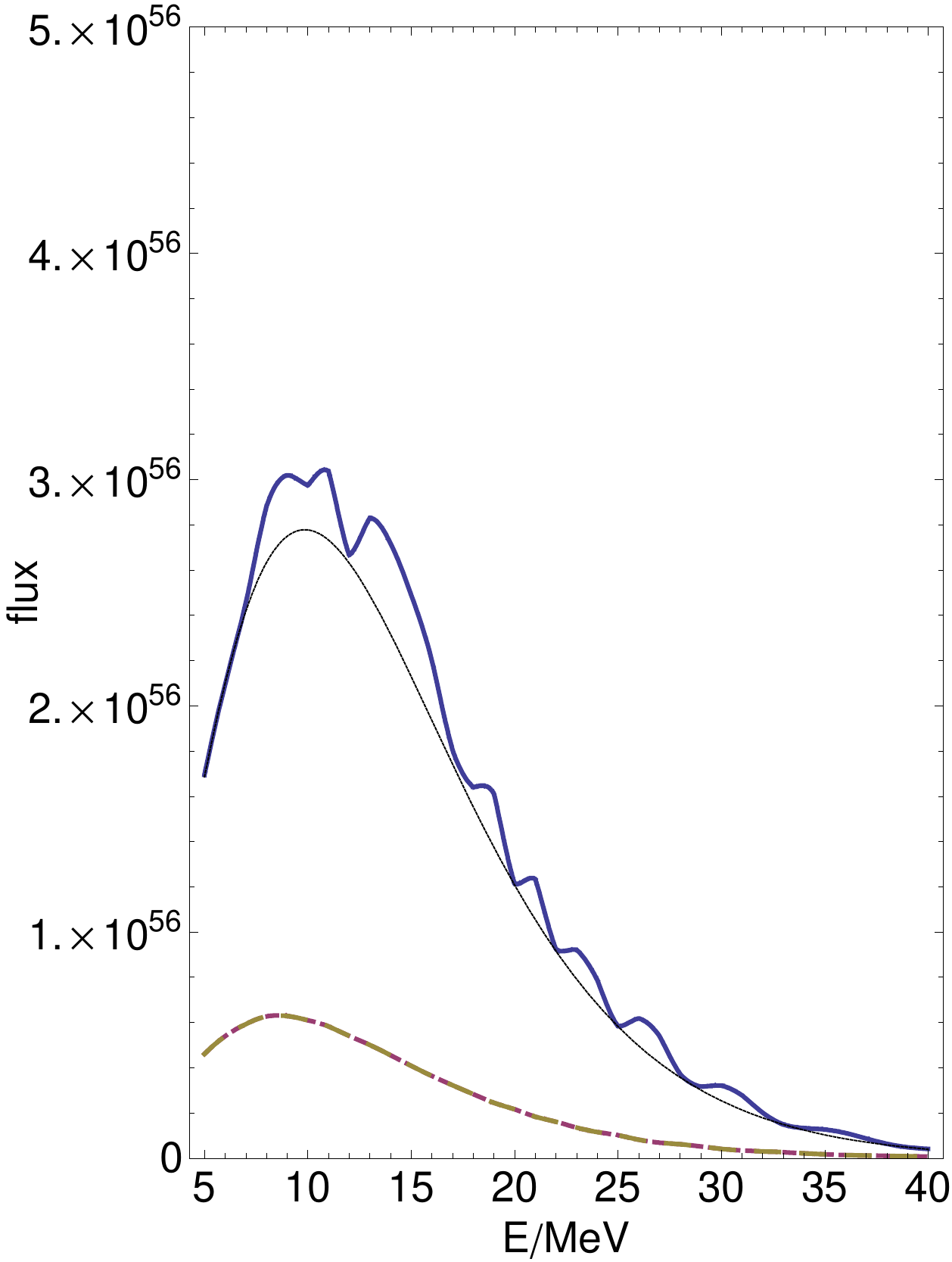}
  \caption{Left column: predicted spectra  of the $\nue$ flux in a detector  for \one\ \sn, inclusive of oscillations in the star and in the Earth (nadir angle 60 degrees). Right column: the same spectra, but with the oscillation effects that are characteristic of a \fe\ \sn.  From the upper to lower panel: $\sin^2\theta_{13}=0.01,6 \cdot 10^{-4}, 10^{-5}$.  The thick curves  refer to different times: $t=60,450,700$ ms (solid, short dashed, and long dashed, respectively).   For $t=60$ ms we also show the spectrum in absence of Earth shielding (thin solid line).  The vertical axis has units of ${\rm MeV^{-1} s^{-1}}$. } 
\label{onespectra}
\end{figure}

\end{widetext}

Our work is focused exclusively on the case of neutrinos and normal mass hierarchy, for two main reasons. The first is that the case of normal hierarchy is particularly simple in its oscillation pattern, and has also been studied in the most detail for \one\ \sne\ in  the pre-shock phase \cite{Duan:2007sh}. This makes it an ideal choice for an initial study of the oscillation signatures of a \one\ \sn.   The second motivation is that the neutrino channel is especially interesting for a \one\ \sn\ because of the very distinctive signature in the neutronization peak, $t \simeq 60 $ ms (fig. \ref{onespectra}).
General arguments on \n\ conversion in the dense matter of a \sn\ (see
e.g. \cite{Dighe:1999bi}) suggest that the features expected for neutrinos and normal hierarchy should appear --
same in character but different in the details -- also in the other cases of  antineutrinos  and/or
inverted mass hierarchy.  Specifically, for the inverted hierarchy all
the phenomena associated with the high density resonance will appear
in the antineutrino channel.  Also, neutrino self-interaction will induce
strong flavor conversion on both \ns\ and anti\ns\ within the first
few hundreds of Km of radius (see e.g. \cite{Duan:2006an}).  

 The generalization of our results to other combinations of channels (\ns\ and anti\ns) and hierarchies require a more in-depth modeling of the effects of \n-\n\
scattering, a difficult task that receives further motivation from this work.  The combination of observations in different channels can only strengthen the conclusions that a single channel can give.

This brings us to the question of what detector is best for the study of the effects we have discussed. 
For the case considered here (neutrino channel), the optimal setup is an experiment that can detect electron neutrinos exclusively. Liquid argon technology is very suitable for this.  To have sufficient statistics in time bins $\sim 100$ ms wide, necessary to test the fast time variations found here, large volumes are required.
Thus, the massive liquid argon project  MODULAr (20 Kt mass, corresponding to $\sim
10^3$ $\nue$ events from a galactic supernova \cite{Bueno:2003ei}) and the even more ambitious
GLACIER \cite{Autiero:2007zj} (up to 100 Kt mass, i.e.  $\sim 6 \cdot
10^3$ $\nue$ events) would be very valuable.  Considering that no large $\nue$ detector is
currently active, after
the closure of SNO \cite{Krauss:2006qq}, it is extremely important to emphasize the
importance of having new detectors of this type to study supernovae,
and our paper contributes in this direction.  Of course, the best out
of a supernova observation would be obtained from combining data from
different detection channels.  The effect on antineutrinos due to the
step in the density of an \one\ \sn\ would be very well visible in the
inverse beta decay events at SuperKamiokande (see e.g.,
\cite{Vagins:2001pi}) and at its larger versions, UNO, HyperKamiokande
and MEMPHYS \cite{Jung:1999jq,Nakamura:2003hk,deBellefon:2006vq}, as
well as at the planned liquid scintillator experiment LENA
\cite{MarrodanUndagoitia:2006re}.  With several detectors running,
there is a substantial chance that at least one of them will be
shielded by the Earth \cite{Lunardini:2001pb,Mirizzi:2006xx} and thus
will have a enhanced sensitivity to \n\ flavor conversion.

CL acknowledges support from Arizona State University, from the RIKEN BNL Research Center (RBRC), and from the ORNL grant of the Institute of Nuclear Theory (INT) of Seattle, where this work was initiated.  The INT is also thanked for hospitality during part of the time of preparation of this work.   CL  is grateful to H.~Duan, A. Marrone and A. Mirizzi for useful clarifications and discussions. 
BM and HTJ are grateful to K.~Nomoto for providing them his stellar
progenitor data and to A.~Marek for his contributions to the
microphysics used in the supernova runs. The project in Garching
was supported by the Deutsche Forschungsgemeinschaft
through the Transregional Collaborative Research Centers SFB/TR~27
``Neutrinos and Beyond'' and SFB/TR~7 ``Gravitational Wave Astronomy'',
and the Cluster of Excellence EXC~153 ``Origin and Structure of the Universe''
({\tt http://www.universe-cluster.de}). The supernova computations were
performed at the High Performance Computing Center Stuttgart (HLRS) under
grant number SuperN/12758.



\begin{thebibliography}{64}
\expandafter\ifx\csname natexlab\endcsname\relax\def\natexlab#1{#1}\fi
\expandafter\ifx\csname bibnamefont\endcsname\relax
  \def\bibnamefont#1{#1}\fi
\expandafter\ifx\csname bibfnamefont\endcsname\relax
  \def\bibfnamefont#1{#1}\fi
\expandafter\ifx\csname citenamefont\endcsname\relax
  \def\citenamefont#1{#1}\fi
\expandafter\ifx\csname url\endcsname\relax
  \def\url#1{\texttt{#1}}\fi
\expandafter\ifx\csname urlprefix\endcsname\relax\def\urlprefix{URL }\fi
\providecommand{\bibinfo}[2]{#2}
\providecommand{\eprint}[2][]{\url{#2}}

\bibitem[{\citenamefont{Schirato and Fuller}(2002)}]{Schirato:2002tg}
\bibinfo{author}{\bibfnamefont{R.~C.} \bibnamefont{Schirato}} \bibnamefont{and}
  \bibinfo{author}{\bibfnamefont{G.~M.} \bibnamefont{Fuller}}
  (\bibinfo{year}{2002}), \eprint{astro-ph/0205390}.

\bibitem[{\citenamefont{Takahashi et~al.}(2003)\citenamefont{Takahashi, Sato,
  Dalhed, and Wilson}}]{Takahashi:2002yj}
\bibinfo{author}{\bibfnamefont{K.}~\bibnamefont{Takahashi}},
  \bibinfo{author}{\bibfnamefont{K.}~\bibnamefont{Sato}},
  \bibinfo{author}{\bibfnamefont{H.~E.} \bibnamefont{Dalhed}},
  \bibnamefont{and} \bibinfo{author}{\bibfnamefont{J.~R.}
  \bibnamefont{Wilson}}, \bibinfo{journal}{Astropart. Phys.}
  \textbf{\bibinfo{volume}{20}}, \bibinfo{pages}{189} (\bibinfo{year}{2003}),
  \eprint{astro-ph/0212195}.

\bibitem[{\citenamefont{Lunardini and Smirnov}(2003)}]{Lunardini:2003eh}
\bibinfo{author}{\bibfnamefont{C.}~\bibnamefont{Lunardini}} \bibnamefont{and}
  \bibinfo{author}{\bibfnamefont{A.~Y.} \bibnamefont{Smirnov}},
  \bibinfo{journal}{JCAP} \textbf{\bibinfo{volume}{0306}}, \bibinfo{pages}{009}
  (\bibinfo{year}{2003}), \eprint{hep-ph/0302033}.

\bibitem[{\citenamefont{Fogli et~al.}(2003)\citenamefont{Fogli, Lisi,
  Montanino, and Mirizzi}}]{Fogli:2003dw}
\bibinfo{author}{\bibfnamefont{G.~L.} \bibnamefont{Fogli}},
  \bibinfo{author}{\bibfnamefont{E.}~\bibnamefont{Lisi}},
  \bibinfo{author}{\bibfnamefont{D.}~\bibnamefont{Montanino}},
  \bibnamefont{and} \bibinfo{author}{\bibfnamefont{A.}~\bibnamefont{Mirizzi}},
  \bibinfo{journal}{Phys. Rev.} \textbf{\bibinfo{volume}{D68}},
  \bibinfo{pages}{033005} (\bibinfo{year}{2003}), \eprint{hep-ph/0304056}.

\bibitem[{\citenamefont{Tomas et~al.}(2004)}]{Tomas:2004gr}
\bibinfo{author}{\bibfnamefont{R.}~\bibnamefont{Tomas}} \bibnamefont{et~al.},
  \bibinfo{journal}{JCAP} \textbf{\bibinfo{volume}{0409}}, \bibinfo{pages}{015}
  (\bibinfo{year}{2004}), \eprint{astro-ph/0407132}.

\bibitem[{\citenamefont{Fogli et~al.}(2005)\citenamefont{Fogli, Lisi, Mirizzi,
  and Montanino}}]{Fogli:2004ff}
\bibinfo{author}{\bibfnamefont{G.~L.} \bibnamefont{Fogli}},
  \bibinfo{author}{\bibfnamefont{E.}~\bibnamefont{Lisi}},
  \bibinfo{author}{\bibfnamefont{A.}~\bibnamefont{Mirizzi}}, \bibnamefont{and}
  \bibinfo{author}{\bibfnamefont{D.}~\bibnamefont{Montanino}},
  \bibinfo{journal}{JCAP} \textbf{\bibinfo{volume}{0504}}, \bibinfo{pages}{002}
  (\bibinfo{year}{2005}), \eprint{hep-ph/0412046}.

\bibitem[{\citenamefont{Duan et~al.}(2007{\natexlab{a}})\citenamefont{Duan,
  Fuller, Carlson, and Qian}}]{Duan:2007sh}
\bibinfo{author}{\bibfnamefont{H.}~\bibnamefont{Duan}},
  \bibinfo{author}{\bibfnamefont{G.~M.} \bibnamefont{Fuller}},
  \bibinfo{author}{\bibfnamefont{J.}~\bibnamefont{Carlson}}, \bibnamefont{and}
  \bibinfo{author}{\bibfnamefont{Y.-Z.} \bibnamefont{Qian}}
  (\bibinfo{year}{2007}{\natexlab{a}}), \eprint{arXiv:0710.1271 [astro-ph]}.

\bibitem[{\citenamefont{Poelarends et~al.}(2007)\citenamefont{Poelarends,
  Herwig, Langer, and Heger}}]{Poelarends:2007ip}
\bibinfo{author}{\bibfnamefont{A.~J.~T.} \bibnamefont{Poelarends}},
  \bibinfo{author}{\bibfnamefont{F.}~\bibnamefont{Herwig}},
  \bibinfo{author}{\bibfnamefont{N.}~\bibnamefont{Langer}}, \bibnamefont{and}
  \bibinfo{author}{\bibfnamefont{A.}~\bibnamefont{Heger}}
  (\bibinfo{year}{2007}), \eprint{arXiv:0705.4643 [astro-ph]}.


\bibitem{Janka:2007yu}
Janka H.-Th., Marek A., Kitaura F.-S.,
in Proceedings of International Conference
Supernova 1987A: 20 Years After: Supernovae and Gamma-Ray Bursters,
Aspen, Colorado, February 19--23, 2007, Eds.~S.~Immler, K.W.~Weiler,
and R. McCray, AIP Conference Proceedings (American Institute of
Physics, New York), Volume 937, pp.~144--154 (2007); arXiv:0706.3056
[astro-ph].


\bibitem{Janka:2007di}
  H.~T.~Janka, B.~Mueller, F.~S.~Kitaura and R.~Buras,
  arXiv:0712.4237 [astro-ph], submitted to Astronomy \& Astrophysics.

\bibitem[{\citenamefont{Wilson and Mayle}(1988)}]{WilsonMayle}
\bibinfo{author}{\bibfnamefont{J.~R.} \bibnamefont{Wilson}} \bibnamefont{and}
  \bibinfo{author}{\bibfnamefont{R.~W.} \bibnamefont{Mayle}},
  \bibinfo{journal}{Astrophys. J.} \textbf{\bibinfo{volume}{334}},
  \bibinfo{pages}{909} (\bibinfo{year}{1988}).

\bibitem[{\citenamefont{Kitaura et~al.}(2006)\citenamefont{Kitaura, Janka, and
  Hillebrandt}}]{Kitaura:2005bt}
\bibinfo{author}{\bibfnamefont{F.~S.} \bibnamefont{Kitaura}},
  \bibinfo{author}{\bibfnamefont{H.-T.} \bibnamefont{Janka}}, \bibnamefont{and}
  \bibinfo{author}{\bibfnamefont{W.}~\bibnamefont{Hillebrandt}},
  \bibinfo{journal}{Astron. Astrophys.} \textbf{\bibinfo{volume}{450}},
  \bibinfo{pages}{345} (\bibinfo{year}{2006}), \eprint{astro-ph/0512065}.

\bibitem[{\citenamefont{Hillebrandt}(1978)}]{hil78}
\bibinfo{author}{\bibfnamefont{W.}~\bibnamefont{Hillebrandt}},
  \bibinfo{journal}{Space Sci. Rev.} \textbf{\bibinfo{volume}{21}},
  \bibinfo{pages}{639} (\bibinfo{year}{1978}).

\bibitem[{\citenamefont{Wheeler et~al.}(1997)\citenamefont{Wheeler, Cowan, and
  Hillebrandt}}]{whecowhi98}
\bibinfo{author}{\bibfnamefont{J.~C.}~\bibnamefont{Wheeler}},
  \bibinfo{author}{\bibfnamefont{J.~J.}~\bibnamefont{Cowan}}, \bibnamefont{and}
  \bibinfo{author}{\bibfnamefont{W.}~\bibnamefont{Hillebrandt}},
  \bibinfo{journal}{Astrophys. J.} \textbf{\bibinfo{volume}{493}},
  \bibinfo{pages}{L101} (\bibinfo{year}{1997}).

\bibitem[{\citenamefont{Sumiyoshi et~al.}(2001)\citenamefont{Sumiyoshi,
  Terasawa, Mathews, Kajino, Yamada, and Suzuki}}]{sumter01}
\bibinfo{author}{\bibfnamefont{K.}~\bibnamefont{Sumiyoshi}},
  \bibinfo{author}{\bibfnamefont{M.}~\bibnamefont{Terasawa}},
  \bibinfo{author}{\bibfnamefont{G.~J.}~\bibnamefont{Mathews}},
  \bibinfo{author}{\bibfnamefont{T.}~\bibnamefont{Kajino}},
  \bibinfo{author}{\bibfnamefont{S.}~\bibnamefont{Yamada}}, \bibnamefont{and}
  \bibinfo{author}{\bibfnamefont{H.}~\bibnamefont{Suzuki}},
  \bibinfo{journal}{Astrophys. J.} \textbf{\bibinfo{volume}{562}},
  \bibinfo{pages}{880} (\bibinfo{year}{2001}).

\bibitem[{\citenamefont{Wanajo et~al.}(2003)\citenamefont{Wanajo, Tamamura,
  Naoki, Nomoto, Yuhri, Beers, and Nozawa}}]{wanom03}
\bibinfo{author}{\bibfnamefont{S.}~\bibnamefont{Wanajo}},
  \bibinfo{author}{\bibfnamefont{M.}~\bibnamefont{Tamamura}},
  \bibinfo{author}{\bibfnamefont{I.}~\bibnamefont{Naoki}},
  \bibinfo{author}{\bibfnamefont{K.}~\bibnamefont{Nomoto}},
  \bibinfo{author}{\bibfnamefont{I.}~\bibnamefont{Yuhri}},
  \bibinfo{author}{\bibfnamefont{T.~C.}~\bibnamefont{Beers}}, \bibnamefont{and}
  \bibinfo{author}{\bibfnamefont{S.}~\bibnamefont{Nozawa}},
  \bibinfo{journal}{Astrophys. J.} \textbf{\bibinfo{volume}{593}},
  \bibinfo{pages}{968} (\bibinfo{year}{2003}).

\bibitem[{\citenamefont{Ning et~al.}(2007)\citenamefont{Ning, Qian,  and
  Meyer}}]{ningqian}
\bibinfo{author}{\bibfnamefont{H.}~\bibnamefont{Ning}},
  \bibinfo{author}{\bibfnamefont{Y.-Z.} \bibnamefont{Qian}}, ,
  \bibnamefont{and} \bibinfo{author}{\bibfnamefont{B.}~\bibnamefont{Meyer}},
  \bibinfo{journal}{Astrophys. J.} \textbf{\bibinfo{volume}{667}},
  \bibinfo{pages}{L159} (\bibinfo{year}{2007}).

\bibitem[{\citenamefont{Gott et~al.}(1970)\citenamefont{Gott, Gunn, and
  Ostriker}}]{gott70}
\bibinfo{author}{\bibfnamefont{J.~R.}~\bibnamefont{Gott}},
  \bibinfo{author}{\bibfnamefont{J.~E.}~\bibnamefont{Gunn}}, \bibnamefont{and}
  \bibinfo{author}{\bibfnamefont{J.~P.} \bibnamefont{Ostriker}},
  \bibinfo{journal}{Astrophys. J.} \textbf{\bibinfo{volume}{160}},
  \bibinfo{pages}{L91} (\bibinfo{year}{1970}).

\bibitem[{\citenamefont{Arnett}(1975)}]{arn75}
\bibinfo{author}{\bibfnamefont{W.}~\bibnamefont{Arnett}},
  \bibinfo{journal}{Astrophys. J.} \textbf{\bibinfo{volume}{195}},
  \bibinfo{pages}{727} (\bibinfo{year}{1975}).

\bibitem[{\citenamefont{Woosley et~al.}(1980)\citenamefont{Woosley, Weaver, and
  Taam}}]{woo80}
\bibinfo{author}{\bibfnamefont{S.}~\bibnamefont{Woosley}},
  \bibinfo{author}{\bibfnamefont{T.}~\bibnamefont{Weaver}}, \bibnamefont{and}
  \bibinfo{author}{\bibfnamefont{R.}~\bibnamefont{Taam}}
  (\bibinfo{year}{1980}), \bibinfo{note}{in {\it Type I Supernovae}, ed. J.C.
  Wheeler (Austin: University of Texas), p.~96}.

\bibitem[{\citenamefont{Hillebrandt}(1982)}]{hil82}
\bibinfo{author}{\bibfnamefont{W.}~\bibnamefont{Hillebrandt}},
  \bibinfo{journal}{Astron. Astrophys.} \textbf{\bibinfo{volume}{110}},
  \bibinfo{pages}{L3} (\bibinfo{year}{1982}).

\bibitem[{\citenamefont{Nomoto et~al.}(1982)\citenamefont{Nomoto, Sparks,
  Fesen, Gull, Miyaji, and Sugimoto}}]{nom82}
\bibinfo{author}{\bibfnamefont{K.}~\bibnamefont{Nomoto}},
  \bibinfo{author}{\bibfnamefont{W.~M.}~\bibnamefont{Sparks}},
  \bibinfo{author}{\bibfnamefont{R.~A.}~\bibnamefont{Fesen}},
  \bibinfo{author}{\bibfnamefont{T.~R.} \bibnamefont{Gull}},
  \bibinfo{author}{\bibfnamefont{S.}~\bibnamefont{Miyaji}}, \bibnamefont{and}
  \bibinfo{author}{\bibfnamefont{D.}~\bibnamefont{Sugimoto}},
  \bibinfo{journal}{Nature} \textbf{\bibinfo{volume}{299}},
  \bibinfo{pages}{803} (\bibinfo{year}{1982}).

\bibitem[{\citenamefont{Mikheev and
  Smirnov}(1986{\natexlab{a}})}]{Mikheev:1986if}
\bibinfo{author}{\bibfnamefont{S.~P.} \bibnamefont{Mikheev}} \bibnamefont{and}
  \bibinfo{author}{\bibfnamefont{A.~Y.} \bibnamefont{Smirnov}},
  \bibinfo{journal}{Sov. Phys. JETP} \textbf{\bibinfo{volume}{64}},
  \bibinfo{pages}{4} (\bibinfo{year}{1986}{\natexlab{a}}),
  \eprint{arXiv:0706.0454 [hep-ph]}.

\bibitem[{\citenamefont{Dighe and Smirnov}(2000)}]{Dighe:1999bi}
\bibinfo{author}{\bibfnamefont{A.~S.} \bibnamefont{Dighe}} \bibnamefont{and}
  \bibinfo{author}{\bibfnamefont{A.~Y.} \bibnamefont{Smirnov}},
  \bibinfo{journal}{Phys. Rev.} \textbf{\bibinfo{volume}{D62}},
  \bibinfo{pages}{033007} (\bibinfo{year}{2000}), \eprint{hep-ph/9907423}.

\bibitem{Dasgupta:2008cd}
  B.~Dasgupta, A.~Dighe, A.~Mirizzi and G.~G.~Raffelt,
  arXiv:0801.1660 [hep-ph].


\bibitem[{\citenamefont{Nomoto}(1984)}]{nom84}
\bibinfo{author}{\bibfnamefont{K.}~\bibnamefont{Nomoto}},
  \bibinfo{journal}{Astrophys. J.} \textbf{\bibinfo{volume}{277}},
  \bibinfo{pages}{791} (\bibinfo{year}{1984}).

\bibitem[{\citenamefont{Nomoto}(1987)}]{nom87}
\bibinfo{author}{\bibfnamefont{K.}~\bibnamefont{Nomoto}},
  \bibinfo{journal}{Astrophys. J.} \textbf{\bibinfo{volume}{322}},
  \bibinfo{pages}{206} (\bibinfo{year}{1987}).

\bibitem[{\citenamefont{Nomoto}()}]{nomotopriv}
\bibinfo{author}{\bibfnamefont{K.}~\bibnamefont{Nomoto}},
  \bibinfo{note}{private communication}.

\bibitem[{\citenamefont{Rampp and Janka}(2002)}]{Rampp:2002bq}
\bibinfo{author}{\bibfnamefont{M.}~\bibnamefont{Rampp}} \bibnamefont{and}
  \bibinfo{author}{\bibfnamefont{H.~T.} \bibnamefont{Janka}},
  \bibinfo{journal}{Astron. Astrophys.} \textbf{\bibinfo{volume}{396}},
  \bibinfo{pages}{361} (\bibinfo{year}{2002}), \eprint{astro-ph/0203101}.

\bibitem[{\citenamefont{Buras et~al.}(2006{\natexlab{a}})\citenamefont{Buras,
  Rampp, Janka, and Kifonidis}}]{Buras:2005rp}
\bibinfo{author}{\bibfnamefont{R.}~\bibnamefont{Buras}},
  \bibinfo{author}{\bibfnamefont{M.}~\bibnamefont{Rampp}},
  \bibinfo{author}{\bibfnamefont{H.~T.} \bibnamefont{Janka}}, \bibnamefont{and}
  \bibinfo{author}{\bibfnamefont{K.}~\bibnamefont{Kifonidis}},
  \bibinfo{journal}{Astron. Astrophys.} \textbf{\bibinfo{volume}{447}},
  \bibinfo{pages}{1049} (\bibinfo{year}{2006}{\natexlab{a}}),
  \eprint{astro-ph/0507135}.


\bibitem[{\citenamefont{Buras et~al.}(2006{\natexlab{b}})\citenamefont{Buras,
  Janka, Rampp, and Kifonidis}}]{Buras:2005tb}
\bibinfo{author}{\bibfnamefont{R.}~\bibnamefont{Buras}},
  \bibinfo{author}{\bibfnamefont{H.-T.} \bibnamefont{Janka}},
  \bibinfo{author}{\bibfnamefont{M.}~\bibnamefont{Rampp}}, \bibnamefont{and}
  \bibinfo{author}{\bibfnamefont{K.}~\bibnamefont{Kifonidis}},
  \bibinfo{journal}{Astron. Astrophys.} \textbf{\bibinfo{volume}{457}},
  \bibinfo{pages}{281} (\bibinfo{year}{2006}{\natexlab{b}}),
  \eprint{astro-ph/0512189}.

\bibitem[{\citenamefont{Keil et~al.}(2003)\citenamefont{Keil, Raffelt, and
  Janka}}]{Keil:2002in}
\bibinfo{author}{\bibfnamefont{M.~T.} \bibnamefont{Keil}},
  \bibinfo{author}{\bibfnamefont{G.~G.} \bibnamefont{Raffelt}},
  \bibnamefont{and} \bibinfo{author}{\bibfnamefont{H.-T.} \bibnamefont{Janka}},
  \bibinfo{journal}{Astrophys. J.} \textbf{\bibinfo{volume}{590}},
  \bibinfo{pages}{971} (\bibinfo{year}{2003}), \eprint{astro-ph/0208035}.

\bibitem[{\citenamefont{Kotake et~al.}(2006)\citenamefont{Kotake, Sato, and
  Takahashi}}]{Kotake:2005zn}
\bibinfo{author}{\bibfnamefont{K.}~\bibnamefont{Kotake}},
  \bibinfo{author}{\bibfnamefont{K.}~\bibnamefont{Sato}}, \bibnamefont{and}
  \bibinfo{author}{\bibfnamefont{K.}~\bibnamefont{Takahashi}},
  \bibinfo{journal}{Rept. Prog. Phys.} \textbf{\bibinfo{volume}{69}},
  \bibinfo{pages}{971} (\bibinfo{year}{2006}), \eprint{astro-ph/0509456}.

\bibitem[{\citenamefont{Fogli et~al.}(2007)\citenamefont{Fogli, Lisi, Marrone,
  and Mirizzi}}]{Fogli:2007bk}
\bibinfo{author}{\bibfnamefont{G.~L.} \bibnamefont{Fogli}},
  \bibinfo{author}{\bibfnamefont{E.}~\bibnamefont{Lisi}},
  \bibinfo{author}{\bibfnamefont{A.}~\bibnamefont{Marrone}}, \bibnamefont{and}
  \bibinfo{author}{\bibfnamefont{A.}~\bibnamefont{Mirizzi}}
  (\bibinfo{year}{2007}), \eprint{arXiv:0707.1998 [hep-ph]}.

\bibitem[{\citenamefont{Gonzalez-Garcia and
  Maltoni}(2007)}]{GonzalezGarcia:2007ib}
\bibinfo{author}{\bibfnamefont{M.~C.} \bibnamefont{Gonzalez-Garcia}}
  \bibnamefont{and} \bibinfo{author}{\bibfnamefont{M.}~\bibnamefont{Maltoni}}
  (\bibinfo{year}{2007}), \eprint{arXiv:0704.1800 [hep-ph]}.

\bibitem[{\citenamefont{Apollonio et~al.}(1999)}]{Apollonio:1999ae}
\bibinfo{author}{\bibfnamefont{M.}~\bibnamefont{Apollonio}}
  \bibnamefont{et~al.} (\bibinfo{collaboration}{CHOOZ}),
  \bibinfo{journal}{Phys. Lett.} \textbf{\bibinfo{volume}{B466}},
  \bibinfo{pages}{415} (\bibinfo{year}{1999}), \eprint{hep-ex/9907037}.

\bibitem[{\citenamefont{Boehm et~al.}(2001)}]{Boehm:2001ik}
\bibinfo{author}{\bibfnamefont{F.}~\bibnamefont{Boehm}} \bibnamefont{et~al.},
  \bibinfo{journal}{Phys. Rev.} \textbf{\bibinfo{volume}{D64}},
  \bibinfo{pages}{112001} (\bibinfo{year}{2001}), \eprint{hep-ex/0107009}.

\bibitem[{\citenamefont{Akhmedov et~al.}(2002)\citenamefont{Akhmedov,
  Lunardini, and Smirnov}}]{Akhmedov:2002zj}
\bibinfo{author}{\bibfnamefont{E.~K.} \bibnamefont{Akhmedov}},
  \bibinfo{author}{\bibfnamefont{C.}~\bibnamefont{Lunardini}},
  \bibnamefont{and} \bibinfo{author}{\bibfnamefont{A.~Y.}
  \bibnamefont{Smirnov}}, \bibinfo{journal}{Nucl. Phys.}
  \textbf{\bibinfo{volume}{B643}}, \bibinfo{pages}{339} (\bibinfo{year}{2002}),
  \eprint{hep-ph/0204091}.

\bibitem[{\citenamefont{Balantekin et~al.}(2007)\citenamefont{Balantekin, Gava,
  and Volpe}}]{Balantekin:2007es}
\bibinfo{author}{\bibfnamefont{A.~B.} \bibnamefont{Balantekin}},
  \bibinfo{author}{\bibfnamefont{J.}~\bibnamefont{Gava}}, \bibnamefont{and}
  \bibinfo{author}{\bibfnamefont{C.}~\bibnamefont{Volpe}}
  (\bibinfo{year}{2007}), \eprint{arXiv:0710.3112 [astro-ph]}.

\bibitem[{\citenamefont{Duan et~al.}(2006{\natexlab{a}})\citenamefont{Duan,
  Fuller, Carlson, and Qian}}]{Duan:2006an}
\bibinfo{author}{\bibfnamefont{H.}~\bibnamefont{Duan}},
  \bibinfo{author}{\bibfnamefont{G.~M.} \bibnamefont{Fuller}},
  \bibinfo{author}{\bibfnamefont{J.}~\bibnamefont{Carlson}}, \bibnamefont{and}
  \bibinfo{author}{\bibfnamefont{Y.-Z.} \bibnamefont{Qian}},
  \bibinfo{journal}{Phys. Rev.} \textbf{\bibinfo{volume}{D74}},
  \bibinfo{pages}{105014} (\bibinfo{year}{2006}{\natexlab{a}}),
  \eprint{astro-ph/0606616}.

\bibitem[{\citenamefont{Duan et~al.}(2006{\natexlab{b}})\citenamefont{Duan,
  Fuller, and Qian}}]{Duan:2005cp}
\bibinfo{author}{\bibfnamefont{H.}~\bibnamefont{Duan}},
  \bibinfo{author}{\bibfnamefont{G.~M.} \bibnamefont{Fuller}},
  \bibnamefont{and} \bibinfo{author}{\bibfnamefont{Y.-Z.} \bibnamefont{Qian}},
  \bibinfo{journal}{Phys. Rev.} \textbf{\bibinfo{volume}{D74}},
  \bibinfo{pages}{123004} (\bibinfo{year}{2006}{\natexlab{b}}),
  \eprint{astro-ph/0511275}.

\bibitem[{\citenamefont{Hannestad et~al.}(2006)\citenamefont{Hannestad,
  Raffelt, Sigl, and Wong}}]{Hannestad:2006nj}
\bibinfo{author}{\bibfnamefont{S.}~\bibnamefont{Hannestad}},
  \bibinfo{author}{\bibfnamefont{G.~G.} \bibnamefont{Raffelt}},
  \bibinfo{author}{\bibfnamefont{G.}~\bibnamefont{Sigl}}, \bibnamefont{and}
  \bibinfo{author}{\bibfnamefont{Y.~Y.~Y.} \bibnamefont{Wong}},
  \bibinfo{journal}{Phys. Rev.} \textbf{\bibinfo{volume}{D74}},
  \bibinfo{pages}{105010} (\bibinfo{year}{2006}), \eprint{astro-ph/0608695}.

\bibitem[{\citenamefont{Duan et~al.}(2007{\natexlab{b}})\citenamefont{Duan,
  Fuller, Carlson, and Qian}}]{Duan:2007mv}
\bibinfo{author}{\bibfnamefont{H.}~\bibnamefont{Duan}},
  \bibinfo{author}{\bibfnamefont{G.~M.} \bibnamefont{Fuller}},
  \bibinfo{author}{\bibfnamefont{J.}~\bibnamefont{Carlson}}, \bibnamefont{and}
  \bibinfo{author}{\bibfnamefont{Y.-Z.} \bibnamefont{Qian}},
  \bibinfo{journal}{Phys. Rev.} \textbf{\bibinfo{volume}{D75}},
  \bibinfo{pages}{125005} (\bibinfo{year}{2007}{\natexlab{b}}),
  \eprint{astro-ph/0703776}.

\bibitem[{\citenamefont{Raffelt and
  Smirnov}(2007{\natexlab{a}})}]{Raffelt:2007cb}
\bibinfo{author}{\bibfnamefont{G.~G.} \bibnamefont{Raffelt}} \bibnamefont{and}
  \bibinfo{author}{\bibfnamefont{A.~Y.} \bibnamefont{Smirnov}}
  (\bibinfo{year}{2007}{\natexlab{a}}), \eprint{arXiv:0705.1830 [hep-ph]}.

\bibitem[{\citenamefont{Duan et~al.}(2007{\natexlab{c}})\citenamefont{Duan,
  Fuller, and Qian}}]{Duan:2007fw}
\bibinfo{author}{\bibfnamefont{H.}~\bibnamefont{Duan}},
  \bibinfo{author}{\bibfnamefont{G.~M.} \bibnamefont{Fuller}},
  \bibnamefont{and} \bibinfo{author}{\bibfnamefont{Y.-Z.} \bibnamefont{Qian}},
  \bibinfo{journal}{Phys. Rev.} \textbf{\bibinfo{volume}{D76}},
  \bibinfo{pages}{085013} (\bibinfo{year}{2007}{\natexlab{c}}),
  \eprint{arXiv:0706.4293 [astro-ph]}.

\bibitem[{\citenamefont{Duan et~al.}(2007{\natexlab{d}})\citenamefont{Duan,
  Fuller, Carlson, and Zhong}}]{Duan:2007bt}
\bibinfo{author}{\bibfnamefont{H.}~\bibnamefont{Duan}},
  \bibinfo{author}{\bibfnamefont{G.~M.} \bibnamefont{Fuller}},
  \bibinfo{author}{\bibfnamefont{J.}~\bibnamefont{Carlson}}, \bibnamefont{and}
  \bibinfo{author}{\bibfnamefont{Y.-Q.} \bibnamefont{Zhong}}
  (\bibinfo{year}{2007}{\natexlab{d}}), \eprint{arXiv:0707.0290 [astro-ph]}.

\bibitem[{\citenamefont{Raffelt and
  Smirnov}(2007{\natexlab{b}})}]{Raffelt:2007xt}
\bibinfo{author}{\bibfnamefont{G.~G.} \bibnamefont{Raffelt}} \bibnamefont{and}
  \bibinfo{author}{\bibfnamefont{A.~Y.} \bibnamefont{Smirnov}}
  (\bibinfo{year}{2007}{\natexlab{b}}), \eprint{arXiv:0709.4641 [hep-ph]}.

\bibitem{Dasgupta:2007ws}
  B.~Dasgupta and A.~Dighe,
  arXiv:0712.3798 [hep-ph].


\bibitem{Duan:2008za}
  H.~Duan, G.~M.~Fuller and Y.~Z.~Qian,
  arXiv:0801.1363 [hep-ph].


  
\bibitem[{\citenamefont{Wolfenstein}(1978)}]{Wolfenstein:1977ue}
\bibinfo{author}{\bibfnamefont{L.}~\bibnamefont{Wolfenstein}},
  \bibinfo{journal}{Phys. Rev.} \textbf{\bibinfo{volume}{D17}},
  \bibinfo{pages}{2369} (\bibinfo{year}{1978}).

\bibitem[{\citenamefont{Mikheev and Smirnov}(1985)}]{Mikheev:1986gs}
\bibinfo{author}{\bibfnamefont{S.~P.} \bibnamefont{Mikheev}} \bibnamefont{and}
  \bibinfo{author}{\bibfnamefont{A.~Y.} \bibnamefont{Smirnov}},
  \bibinfo{journal}{Sov. J. Nucl. Phys.} \textbf{\bibinfo{volume}{42}},
  \bibinfo{pages}{913} (\bibinfo{year}{1985}).

\bibitem[{\citenamefont{Mikheev and
  Smirnov}(1986{\natexlab{b}})}]{Mikheev:1986wj}
\bibinfo{author}{\bibfnamefont{S.~P.} \bibnamefont{Mikheev}} \bibnamefont{and}
  \bibinfo{author}{\bibfnamefont{A.~Y.} \bibnamefont{Smirnov}},
  \bibinfo{journal}{Nuovo Cim.} \textbf{\bibinfo{volume}{C9}},
  \bibinfo{pages}{17} (\bibinfo{year}{1986}{\natexlab{b}}).

\bibitem[{\citenamefont{Fogli et~al.}(2002)\citenamefont{Fogli, Lisi,
  Montanino, and Palazzo}}]{Fogli:2001pm}
\bibinfo{author}{\bibfnamefont{G.~L.} \bibnamefont{Fogli}},
  \bibinfo{author}{\bibfnamefont{E.}~\bibnamefont{Lisi}},
  \bibinfo{author}{\bibfnamefont{D.}~\bibnamefont{Montanino}},
  \bibnamefont{and} \bibinfo{author}{\bibfnamefont{A.}~\bibnamefont{Palazzo}},
  \bibinfo{journal}{Phys. Rev.} \textbf{\bibinfo{volume}{D65}},
  \bibinfo{pages}{073008} (\bibinfo{year}{2002}), \eprint{hep-ph/0111199}.


\bibitem[{\citenamefont{Duan}()}]{duanpriv}
\bibinfo{author}{\bibfnamefont{H.}~\bibnamefont{Duan}}, \bibinfo{note}{private
  communication}.

\bibitem[{\citenamefont{Nunokawa et~al.}(1997)\citenamefont{Nunokawa, Semikoz,
  Smirnov, and Valle}}]{Nunokawa:1997dp}
\bibinfo{author}{\bibfnamefont{H.}~\bibnamefont{Nunokawa}},
  \bibinfo{author}{\bibfnamefont{V.~B.} \bibnamefont{Semikoz}},
  \bibinfo{author}{\bibfnamefont{A.~Y.} \bibnamefont{Smirnov}},
  \bibnamefont{and} \bibinfo{author}{\bibfnamefont{J.~W.~F.}
  \bibnamefont{Valle}}, \bibinfo{journal}{Nucl. Phys.}
  \textbf{\bibinfo{volume}{B501}}, \bibinfo{pages}{17} (\bibinfo{year}{1997}),
  \eprint{hep-ph/9701420}.

\bibitem[{\citenamefont{Lunardini and Smirnov}(2001)}]{Lunardini:2001pb}
\bibinfo{author}{\bibfnamefont{C.}~\bibnamefont{Lunardini}} \bibnamefont{and}
  \bibinfo{author}{\bibfnamefont{A.~Y.} \bibnamefont{Smirnov}},
  \bibinfo{journal}{Nucl. Phys.} \textbf{\bibinfo{volume}{B616}},
  \bibinfo{pages}{307} (\bibinfo{year}{2001}), \eprint{hep-ph/0106149}.

\bibitem[{\citenamefont{Dighe et~al.}(2003)\citenamefont{Dighe, Keil, and
  Raffelt}}]{Dighe:2003jg}
\bibinfo{author}{\bibfnamefont{A.~S.} \bibnamefont{Dighe}},
  \bibinfo{author}{\bibfnamefont{M.~T.} \bibnamefont{Keil}}, \bibnamefont{and}
  \bibinfo{author}{\bibfnamefont{G.~G.} \bibnamefont{Raffelt}},
  \bibinfo{journal}{JCAP} \textbf{\bibinfo{volume}{0306}}, \bibinfo{pages}{006}
  (\bibinfo{year}{2003}), \eprint{hep-ph/0304150}.

\bibitem[{\citenamefont{Dighe et~al.}(2004)\citenamefont{Dighe, Kachelriess,
  Raffelt, and Tomas}}]{Dighe:2003vm}
\bibinfo{author}{\bibfnamefont{A.~S.} \bibnamefont{Dighe}},
  \bibinfo{author}{\bibfnamefont{M.}~\bibnamefont{Kachelriess}},
  \bibinfo{author}{\bibfnamefont{G.~G.} \bibnamefont{Raffelt}},
  \bibnamefont{and} \bibinfo{author}{\bibfnamefont{R.}~\bibnamefont{Tomas}},
  \bibinfo{journal}{JCAP} \textbf{\bibinfo{volume}{0401}}, \bibinfo{pages}{004}
  (\bibinfo{year}{2004}), \eprint{hep-ph/0311172}.

\bibitem[{\citenamefont{Bueno et~al.}(2003)\citenamefont{Bueno, Gil-Botella,
  and Rubbia}}]{Bueno:2003ei}
\bibinfo{author}{\bibfnamefont{A.}~\bibnamefont{Bueno}},
  \bibinfo{author}{\bibfnamefont{I.}~\bibnamefont{Gil-Botella}},
  \bibnamefont{and} \bibinfo{author}{\bibfnamefont{A.}~\bibnamefont{Rubbia}}
  (\bibinfo{year}{2003}), \eprint{hep-ph/0307222}.

\bibitem[{\citenamefont{Autiero et~al.}(2007)}]{Autiero:2007zj}
\bibinfo{author}{\bibfnamefont{D.}~\bibnamefont{Autiero}} \bibnamefont{et~al.},
  \bibinfo{journal}{JCAP} \textbf{\bibinfo{volume}{0711}}, \bibinfo{pages}{011}
  (\bibinfo{year}{2007}), \eprint{arXiv:0705.0116 [hep-ph]}.

\bibitem[{\citenamefont{Krauss}(2006)}]{Krauss:2006qq}
\bibinfo{author}{\bibfnamefont{C.~B.} \bibnamefont{Krauss}}
  (\bibinfo{collaboration}{SNO}), \bibinfo{journal}{J. Phys. Conf. Ser.}
  \textbf{\bibinfo{volume}{39}}, \bibinfo{pages}{275} (\bibinfo{year}{2006}).

\bibitem[{\citenamefont{Vagins}(2001)}]{Vagins:2001pi}
\bibinfo{author}{\bibfnamefont{M.~R.} \bibnamefont{Vagins}},
  \bibinfo{journal}{Int. J. Mod. Phys.} \textbf{\bibinfo{volume}{A16S1B}},
  \bibinfo{pages}{724} (\bibinfo{year}{2001}).

\bibitem[{\citenamefont{Jung}(1999)}]{Jung:1999jq}
\bibinfo{author}{\bibfnamefont{C.~K.} \bibnamefont{Jung}}
  (\bibinfo{year}{1999}), \eprint{hep-ex/0005046}.

\bibitem[{\citenamefont{Nakamura}(2003)}]{Nakamura:2003hk}
\bibinfo{author}{\bibfnamefont{K.}~\bibnamefont{Nakamura}},
  \bibinfo{journal}{Int. J. Mod. Phys.} \textbf{\bibinfo{volume}{A18}},
  \bibinfo{pages}{4053} (\bibinfo{year}{2003}).

\bibitem[{\citenamefont{de~Bellefon et~al.}(2006)}]{deBellefon:2006vq}
\bibinfo{author}{\bibfnamefont{A.}~\bibnamefont{de~Bellefon}}
  \bibnamefont{et~al.} (\bibinfo{year}{2006}), \eprint{hep-ex/0607026}.

\bibitem[{\citenamefont{Marrodan~Undagoitia
  et~al.}(2006)}]{MarrodanUndagoitia:2006re}
\bibinfo{author}{\bibfnamefont{T.}~\bibnamefont{Marrodan~Undagoitia}}
  \bibnamefont{et~al.}, \bibinfo{journal}{Prog. Part. Nucl. Phys.}
  \textbf{\bibinfo{volume}{57}}, \bibinfo{pages}{283} (\bibinfo{year}{2006}),
  \eprint{hep-ph/0605229}.

\bibitem[{\citenamefont{Mirizzi et~al.}(2006)\citenamefont{Mirizzi, Raffelt,
  and Serpico}}]{Mirizzi:2006xx}
\bibinfo{author}{\bibfnamefont{A.}~\bibnamefont{Mirizzi}},
  \bibinfo{author}{\bibfnamefont{G.~G.} \bibnamefont{Raffelt}},
  \bibnamefont{and} \bibinfo{author}{\bibfnamefont{P.~D.}
  \bibnamefont{Serpico}}, \bibinfo{journal}{JCAP}
  \textbf{\bibinfo{volume}{0605}}, \bibinfo{pages}{012} (\bibinfo{year}{2006}),
  \eprint{astro-ph/0604300}.

\end{thebibliography}

\end{document}